\documentclass[prb,aps,epsf,showpacs,twocolumn]{revtex4-1}
\usepackage{times}
\usepackage{graphicx}
\usepackage{float}
\usepackage{latexsym,amsmath,amssymb,bm,euscript}
\usepackage{color}
\usepackage{subfigure}
\usepackage{epstopdf}
\usepackage[colorlinks=true,linkcolor=blue,citecolor=blue,urlcolor=blue]{hyperref}
\usepackage{hyperref}
\usepackage{ulem}
\usepackage[dvipsnames]{xcolor}

\begin{document}

\title{Nematic spin liquid phase in a frustrated spin-$1$ system on the square lattice}
\author{Wen-Jun Hu$^{1}$} 
\author{Shou-Shu Gong$^{2}$}
\email{shoushu.gong@buaa.edu.cn}
\author{Hsin-Hua Lai$^{1}$}
\author{Haoyu Hu$^{1}$}
\author{Qimiao Si$^{1}$}
\author{Andriy H. Nevidomskyy$^{1}$}
\affiliation{
$^1$ Department of Physics and Astronomy \& Rice Center for Quantum Materials, Rice University, Houston, Texas 77005, USA\\
$^2$ Department of Physics, Beihang University, Beijing 100191, China
}

\begin{abstract}
Frustration in quantum spin systems promote a variety of novel quantum phases. An important example is the frustrated spin-$1$ model on the square lattice with the nearest-neighbor bilinear ($J_1$) and biquadratic ($K_1$) interactions. We provide strong evidence for a nematic spin liquid phase in a range of $K_1/J_1$ near the SU(3)-symmetric point ($J_1 = K_1$), based on the linear flavor-wave theory and extensive density matrix renormalization group calculation. This phase displays no spin dipolar or quadrupolar order, preserves translational symmetry but spontaneously breaks $C_4$ lattice rotational symmetry, and possesses fluctuations peaked at the wavevector $(\pi, 2\pi/3)$. The spin excitation gap drops rapidly with system size and appears to be gapless, and the nematic order is attributed to the dominant $(\pi, 2\pi/3)$ fluctuations. Our results provide a novel mechanism for electronic nematic order and, more generally, open up a new avenue to explore frustration-induced exotic ground states.
\end{abstract}

\pacs{73.43.Nq, 75.10.Jm, 75.10.Kt}
\maketitle

\section{Introduction} 
The spin-$1/2$ Heisenberg models represent a prototype for quantum magnetism. Such models can realize exotic phases of matter if frustration becomes dominant. One of the intriguing phases is the elusive quantum spin liquid (QSL)~\cite{Anderson1973,savary2016}. Stimulated by experimental findings of spin-liquid-like antiferromagnets, frustrated spin-$1/2$ Heisenberg models have been studied extensively and QSLs have been found in realistic models (see Ref.~\onlinecite{savary2016} and references therein).

Spin-$1$ systems are also of extensive interest in the search of novel quantum phases~\cite{Haldane1983_2, aklt1987, katsumata1989, hagiwara1990, white1993, schollwock1996, shelton1996, lauchli2006, corboz2007, corboz2017}. Different phases in spin-$1$ systems have been considered to understand the exotic magnetism and nematic order of the iron-chalcogenide superconductor~\cite{rong2015,wang2016,gong2017,lai2017}. The QSLs with high symmetries have also been proposed in the cold atom systems with multiple flavors of fermions and a large Hubbard interaction~\cite{gorshkov2010, taie2010, beverland2016}. In particular, the non-Abelian chiral spin liquid states, which can realize topological quantum computation~\cite{kitaev2006}, have been sought for a long time in different spin-$1$ models~\cite{martin2014, liu2018, chen2018}. In experiment, spin-liquid-like behaviors have been reported in several spin-$1$ compounds, including triangular (NiGa$_2$S$_4$~\cite{nakatsuji2005}, Ba$_3$NiSb$_2$O$_9$~\cite{cheng2011, fak2017}) and honeycomb (6HB-Ba$_3$NiSb$_2$O$_9$~\cite{quilliam2016}) antiferromagnets. Theoretical and experimental discoveries motivate the search of QSL in spin-$1$ systems. However, since quantum fluctuations are reduced for larger spin and it seems hard for geometric frustration alone to destroy magnetic order, new ingredients are highly desired.

For spin-$1$ (or higher) systems, biquadratic interactions are also allowed in addition to the bilinear ones. The competition between the two types of interactions represents an {\it added} form of frustration, which is particularly strong when the two interactions are comparable. If they have the same magnitude, the system has an enlarged SU(3) symmetry~\cite{papanicolaou1988, toth2012}. This symmetry gives rise to a large classical degeneracy of different magnetic dipolar and quadrupolar orders, which may lead to novel phases when quantum fluctuations are incorporated~\cite{hermele2009}. One of the outstanding questions is whether proximity to an SU(3) symmetry promotes any spin liquid phase. We remark that, to date, evidence for spin liquid states in spin-$1$ models has been scarce~\cite{gong2017} and the model studies incorporating the SU(3) symmetry have only found conventional orders~\cite{lauchli2006, toth2010, toth2012, bauer2012, corboz2012, zhao2012, corboz2013}.

In this paper, we study the quantum phases near the SU(3) point ($J_1=K_1$) of the spin-$1$ bilinear-biquadratic model on the square lattice, which has the Hamiltonian
\begin{equation}\label{model}
H = \sum_{\langle i,j\rangle} J_1 {\bf S}_i \cdot {\bf S}_j + K_1 ( {\bf S}_i \cdot {\bf S}_j )^2 .
\end{equation}
Here, ${\bf S}_{i}$ is the spin-$1$ operator at site $i$, and $J_1,K_1$ are the nearest-neighbor bilinear and biquadratic interactions. We use linear flavor-wave theory (LFWT) to show that, over a range of parameters including the SU(3) point, a $(\pi, 2\pi/3)$ nematic and antiferroquadrupolar (AFQ23) order is energetically competitive, \textit{albeit} ultimately unstable. Then we carry out density matrix renormalization group (DMRG) studies on cylinder geometry with circumference up to $9$. Surprisingly, we identify the system near the SU(3) point as a spin liquid phase without any spin dipolar, quadrupolar, or valence-bond crystal (VBC) order. Equally important, this phase is found to possess a lattice nematic order with spontaneously broken $C_4$ rotational symmetry. We study the low-lying excitations of this nematic liquid phase in different aspects, which suggest possible gapless nature of the spin triplet excitation spectrum and lead to a qualitative understanding for the dominating $(\pi, 2\pi/3)$ fluctuations and, by extension, the origin of the nematic order.

It should be noted that the square-lattice model defined by Eq.~\eqref{model} contains a delicate interplay between the three-flavor spin degrees of freedom and the bipartite nature of the lattice~\cite{toth2012}. Previous semi-classical studies using a site-factorized wavefunction~\cite{papanicolaou1988, toth2012} found the SU(3) point as the boundary between the N\'eel AFM phase (AFM2) and a ``semiordered phase'' with infinitely degenerate ground states. For the SU(3) model, the inclusion of quantum fluctuations at the level of LFWT reported a ground state with three-sublattice magnetic order (AFM3) at momentum ${\bf q} = (2\pi/3, 2\pi/3)$~\cite{toth2010, toth2012}; such an order has also been suggested by numerical calculations~\cite{toth2010, toth2012, bauer2012}. Nonetheless, both the LFWT~\cite{toth2010} and the Schwinger boson mean-field~\cite{bauer2012} calculations had difficulties in obtaining a finite value of the AFM3 order parameter, which suggests that strong fluctuations may result in a novel phase beyond the grasp of mean-field or flavor-wave theories.

This paper is organized as follow.
In Sec.~\ref{sec:theory}, we extend the LFWT to study more possible magnetic dipolar and quadrupolar orders.
We emphasize a possible nematic order from the fluctuations of the AFQ23 order.
In Sec.~\ref{sec:dmrg}, we show our extensive DMRG results, which characterize the intermediate phase near the SU(3) point as a nematic spin liquid.
The large-size data strongly support the absent magnetic dipolar and quadrupolar order, but the existence of a spontaneous lattice rotational symmetry breaking.
In Sec.~\ref{sec:discussion}, we further discuss the nature of the intermediate phase.
Our results exclude the existence of a Haldane-like phase but suggest a gapless spin liquid state with emergent Fermi points.
In Sec.~\ref{sec:afm3}, we show the DMRG phase diagram of the SU(3) model with additional third-neighbor biquadratic interaction $K_3$, which reveals a region of the AFM3 phase, which has been suggested as the ground state of the SU(3) model.
Our results are summarized in Sec.~\ref{sec:summary}. 
 
\section{Flavor wave theory calculation}
\label{sec:theory}

First of all, we use LFWT to study the vicinity of the SU(3) point and consider different competing orders with finite stiffness. 
The similar calculations have been done for the AFM3 and the AFQ3 order~\cite{bauer2012}.
Inspired by DMRG results, we explore more possible states here.

\subsection{General formalism of the flavor wave theory}

Depending on the nature of the orders that we consider (either a magnetic or quadrupolar order), we choose either the time-reversal invariant basis of the SU(3) fundamental representation or the usual spin $S^z$ basis, respectively, which can be related by
\begin{equation}\label{supp:lfwt_basis}
\begin{array}{ccc}
|x\rangle = \frac{ i | 1 \rangle - i | \bar{1}\rangle}{\sqrt{2}},&~|y\rangle = \frac{| 1\rangle + | \bar{1}\rangle}{\sqrt{2}},&~ | z \rangle = -i | 0 \rangle,
\end{array}
\end{equation}
where we abbreviate $|S^z = \pm1\rangle \equiv |\pm1\rangle$ $(|S^z = 0\rangle \equiv |0\rangle)$ and $|\bar{1}\rangle \equiv |-1\rangle$. 
The $|x \rangle$, $|y \rangle$ and $|z \rangle$ are the time-reversal invariant basis and are more convenient for performing the flavor wave theory calculations for the quadrupolar orders, while the other basis in $S^z$ are more suitable for the magnetic orders. 
For the flavor wave theory, we associate $3$ Schwinger-bosons at each site $i$, $b_{i\alpha}$, to the states of Eq.~\eqref{supp:lfwt_basis}, where $b^\dagger_{i\alpha} |vac\rangle = |\alpha\rangle$ with $|vac\rangle$ being the vacuum state of the Schwinger bosons and $\alpha = x,~y,~z$ or $ 0, 1, \bar{1}$ depending on the nature of the orders that we consider. 
The bosons satisfy a local constraint $\sum_{\alpha} b^\dagger_{i \alpha}b_{i \alpha} = 1$.
Therefore, the model Hamiltonian Eq.~\eqref{model} can be rewritten as
\begin{eqnarray}\label{Eq:Boson_H}
H = \sum_{i, \delta_n, \alpha, \beta} \left[ J_n b^\dagger_{i \alpha} b_{j \alpha} b^\dagger_{j\beta} b_{i\beta} + \left(K_n - J_n\right) b_{i\alpha}^\dagger b^\dagger_{j \alpha} b_{j\beta} b_{i\beta}\right]. \nonumber
\end{eqnarray}
Following the usual procedure of the spin-wave theory calculation, we introduce different local rotations for each sublattice in different orders. For AFQ orders, we introduce
\begin{eqnarray}
\begin{pmatrix}
a_{ix} \\
a_{iy} \\
a_{iz}
\end{pmatrix} =
\begin{pmatrix}
\cos\theta_i & \sin\theta_i & 0 \\
-\sin\theta_i & \cos\theta_i & 0\\
0 & 0 & 1
\end{pmatrix}
\begin{pmatrix}
b_{ix}\\
b_{iy}\\
b_{iz}
\end{pmatrix},
\end{eqnarray}
which preserve the local constraint with $b_{i\alpha} \rightarrow a_{i\alpha}$. We assume that at each site only one flavor of bosons $a_{ix}$ condenses, and we replace $a^\dagger_{ix}$ and $a_{ix}$ by $(M - a_{iy}^\dagger a_{iy} - a^\dagger_{iz} a_{iz})^{\frac{1}{2}}$, with $M=1$ in the present case. 
A $1/M$ expansion up to the quadratic order of the  bosons $a_y$ and $a_z$ followed by an appropriate Holstein-Primakoff transformation allows us to extract the ground state energy.
For magnetic orders, we make local rotations at each site $i$ to align each spin at different sites along $|1\rangle$, and replace $a^\dagger_{i 1}$ and $a_{i 1}$ by $(M - a_{i 0}^\dagger - a^\dagger_{i \bar{1}})^{\frac{1}{2}}$, and follow the same procedure above to perform a $1/M$ expansion up to the quadratic oder of the bosons $a_{0}$ and $a_{\bar{1}}$. 

\begin{figure}[t]
\includegraphics[width = 3 in]{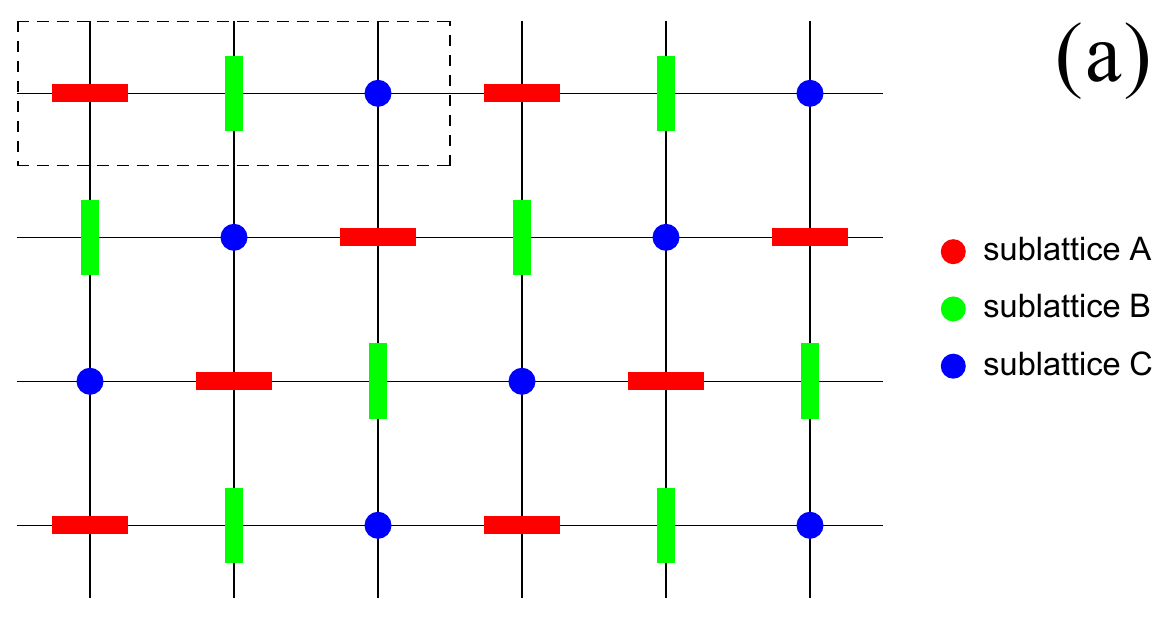}
\includegraphics[width = 3 in]{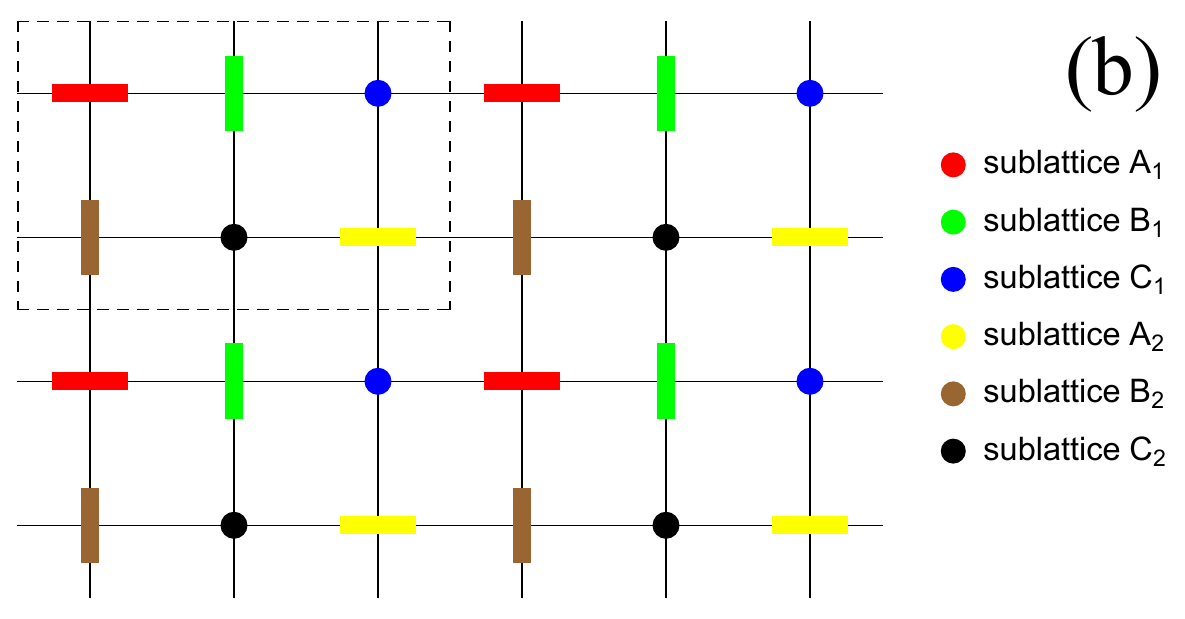}
\caption{Illustration of (a) the AFQ3 order and (b) the AFQ23 order. 
In both figures, the different sublattices are denoted by different colors.
The dashed box includes the sites in an unit cell.}
\label{fig_order}
\end{figure}

\subsection{Results of the flavor-wave theory calculation}

Within the flavor-wave calculation up to the quadrupolar orders, we consider the $(2\pi/3, 2\pi/3)$ AFQ (AFQ3) order, which is a three-sublattice order as shown in Fig.~\ref{fig_order}(a), where the red, green, and blue bars represent the sites occupied by only $b_x$, $b_y$, $b_z$ bosons. 
We also consider the AFQ23 order, whose unit cell contains the $2 \times3$ sublattice structure as shown in Fig.~\ref{fig_order}(b).
This state has been ignored in previous study~\cite{bauer2012}.
The consideration of the AFQ23 order is motivated by the DMRG results, which find the dominant structure factor at $(2\pi/3, \pi)$, instead of $(2\pi/3, 2\pi/3)$, on the finite-size system near the SU(3) model. 
We remind that this AFQ23 order eventually is suppressed at the thermodynamic limit. 
For the magnetic order, we consider the ferromagnetic order (FM) and the N\'eel  AFM order (AFM2). 

\begin{figure}[t]
\includegraphics[width = 3 in ]{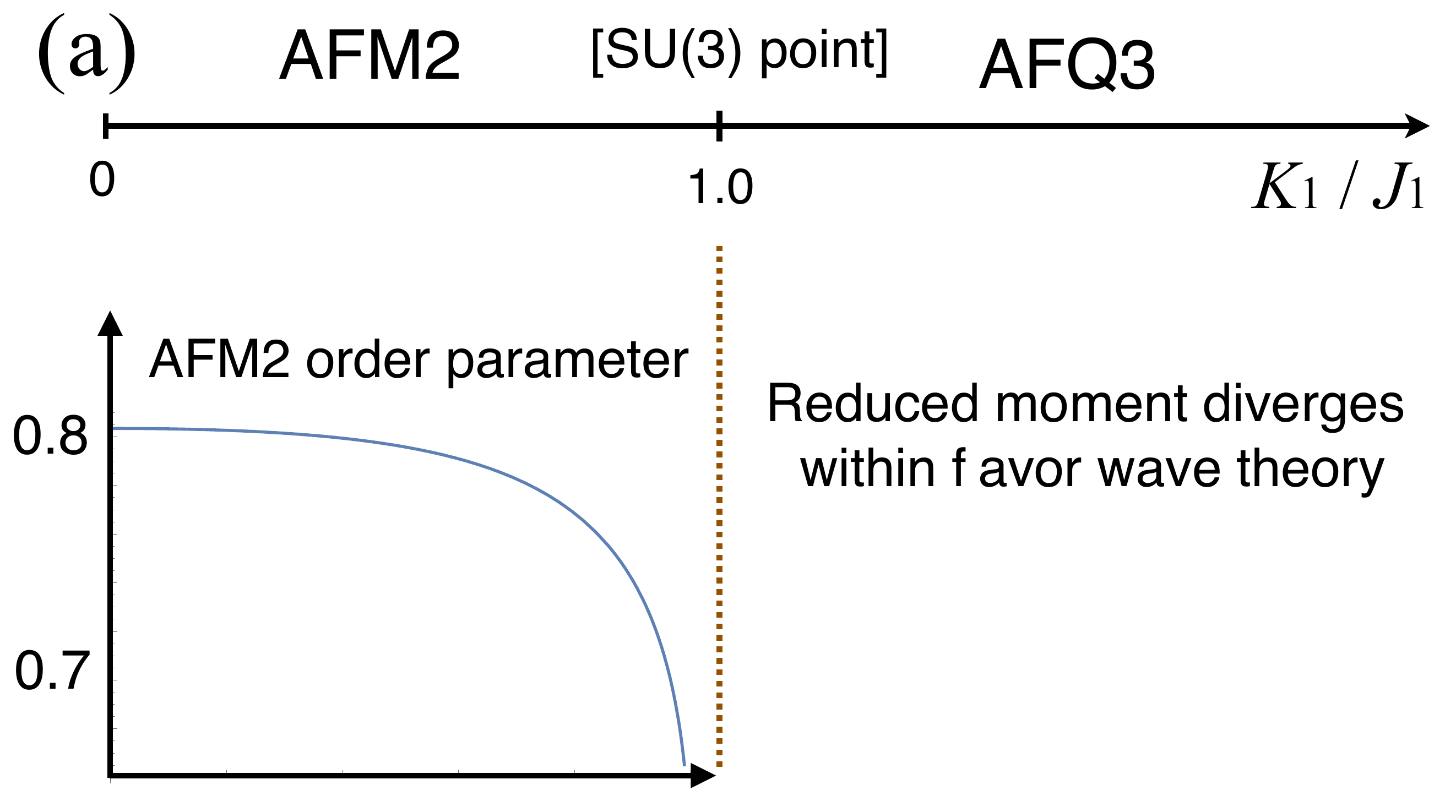}
\includegraphics[width = 2.5 in ]{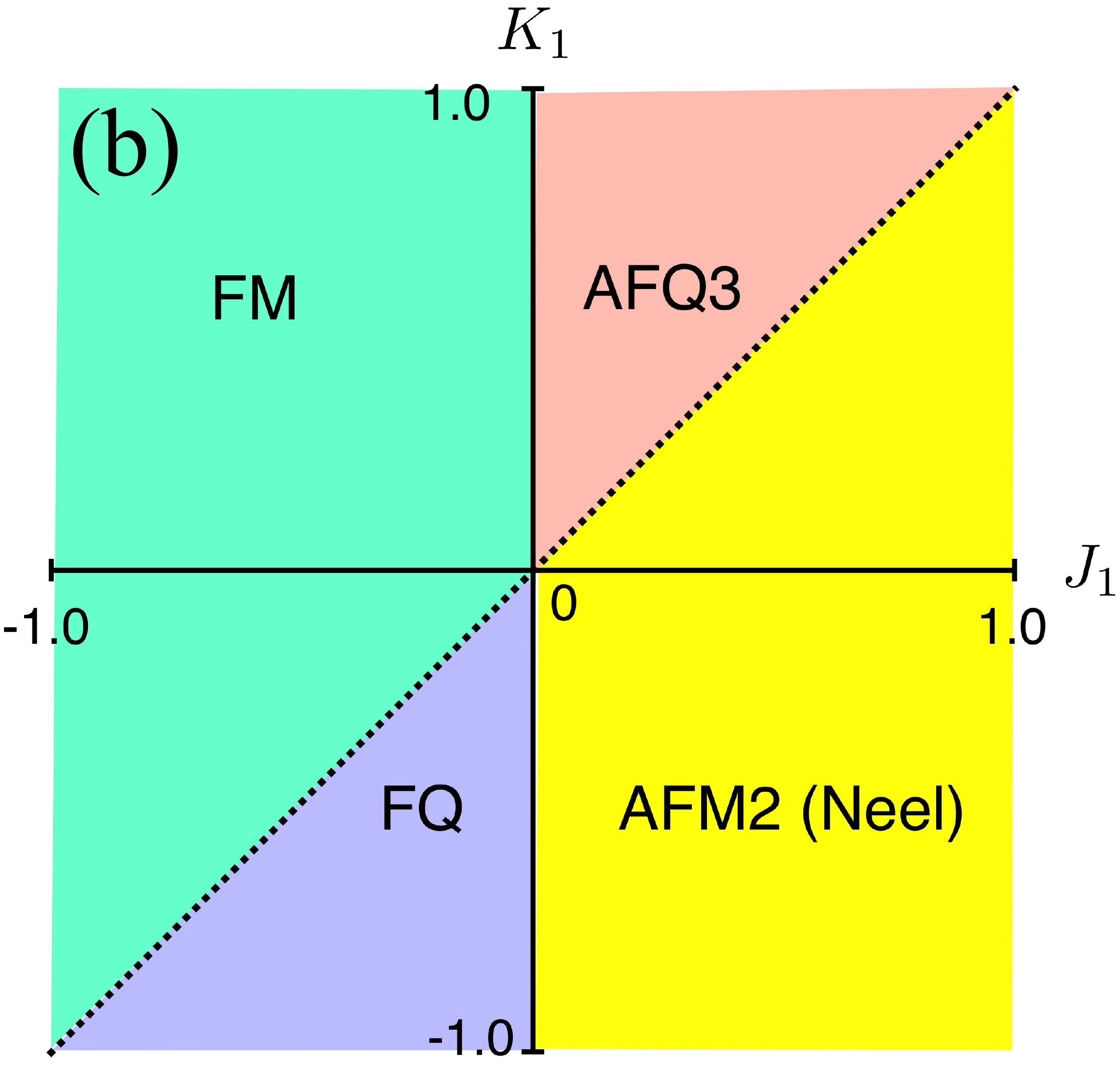}
\caption{Phase diagrams of the spin-$1$ bilinear-biquadratic model Eq.~\eqref{model} obtained based on the energetics within linear flavor-wave theory.
(a) The upper panel is the phase diagram with $J_1 > 0, K_1 > 0$. 
Bottom panel gives the reduction of the order parameter approaching the phase boundary. We remark that the reduction of the AFQ3 moment always diverges indicating that the linear flavor wave calculations in that regime may not be reliable.
(b) The phase diagram based on the flavor-wave theory calculation for different couplings.  The FM refers to the ferromagnetic phase, AFQ3 refers to the AFQ phase ordered at $\bm{q} = (2\pi/3, 2\pi/3)$, FQ refers to the ferroquadrupolar phase with the ordering at $\bm{q} = (0,0)$, and AFM2 corresponds to the N\'eel AFM order. 
The dashed line is the SU(3) line, which is precisely the boundary between AFQ3 and AFM2 and that between FQ and FM.}
\label{fig_supp_phase}
\end{figure}

We leave the details of the calculations to Appendix~\ref{app1} and discuss the obtained results here.
For $J_1, K_1 >0$, the phase diagram based on the energetics is illustrated in the upper panel in Fig.~\ref{fig_supp_phase}(a). 
The result based on the energetics is qualitatively consistent with the previous exact diagonalization result with a small system size (up to 20 sites)~\cite{toth2010, toth2012}. 
Furthermore, we calculate the order parameters for the different orders, i.e. on-site magnetization $\langle {\bf S} \rangle$ for the AFM2 order and the $z$-boson density $\langle a^\dagger_z a_z \rangle = M - \sum_{\alpha = x,y} \langle a^\dagger_{\alpha} a_\alpha \rangle$ for the AFQ3 order, which dictate the reduction of the ordered moments due to the quantum fluctuations.
As illustrated in the bottom panel in Fig.~\ref{fig_supp_phase}(a), for the AFM2 we find that the order parameter is gradually reduced toward the SU(3) point. 
For $K_1/J_1 \geq 1$, we find that for both AFQ3 and AFQ23 orders the reduction of the order moment is always \textit{divergent} due to the presence of gapless lines in the boson dispersions. 
The divergence of the reduction of the order moment was also found previously at the SU(3) point within the flavor-wave theory calculations~\cite{bauer2012}, and here we illustrate that the divergence for AFQ3 not only occurs at the SU(3) point but in the whole regime for $K_1/J_1 \geq 1$. 
For a general case of $J_1$ and $K_1$ taking both signs, we obtain the phase diagram in Fig.~\ref{fig_supp_phase}(b). 
We remark that the phase diagrams in Fig.~\ref{fig_supp_phase} solely based on the energetics within the flavor-wave theory are qualitatively consistent with the previous results~\cite{toth2012}.

\subsection{Nematic order: indications from flavor wave theory}

We note that, in addition to the previously studied AFM3 and AFQ3 orders~\cite{toth2010, toth2012}, the AFM23 and AFQ23 orders at ${\bf q} = (\pi, 2\pi/3)/(2\pi/3, \pi)$ can {\it also} be locally stabilized in our calculation. 
In particular, for $K_1/J_1 \ge 1$, the AFQ23 order is degenerate at the mean-field level with the AFQ3 order; it has higher energy than the latter only as a result of zero-point fluctuations. 
The AFQ23 order is of particular interest because it is accompanied by a nematic order. 
As we have mentioned that the corrections to the quadrupolar order parameters by quantum fluctuations are divergent in the whole parameter regime including the SU(3) point. 
Nonetheless, the quantum correction to the nematic order is finite. 
What emerges is that the static orderings of both AFQ23 and AFQ3 are destroyed by quantum fluctuations but, if the fluctuating order of the AFQ23 type remains important, a static nematic order can still be stabilized, as we will demonstrate using DMRG.

\section{Large-scale DMRG simulation}
\label{sec:dmrg}

\subsection{DMRG phase diagram}

We establish quantum phases based on DMRG~\cite{white1992, mcculloch2002} calculations on cylindrical geometry by keeping up to $4000$ SU(2) states with truncation error below $1\times 10^{-5}$ (see more details in Table~\ref{table} of Appendix~\ref{app2}). We mainly study the rectangular cylinder (RC) with the periodic boundary conditions in the $y$ direction and the open boundaries in the $x$ direction, which are denoted as RC$L_y$-$L_x$ with $L_y$ up to $9$ and $L_x$ up to $36$ ($L_y$ and $L_x$ are the number of sites in the two directions). We have also calculated the RC12-36 cylinder for the SU(3) model by keeping up to $8000$ SU(2) states. The results are consistent with those on the RC6 and RC9 cylinders. In our calculation, we set $J_1 = 1$ as the energy scale. Since DMRG calculation may be strongly affected by finite-size effects, we have also studied other geometries and boundary conditions (see results and discussion in Appendix~\ref{app2}).

\begin{figure}[t]
\includegraphics[width=1.0\linewidth]{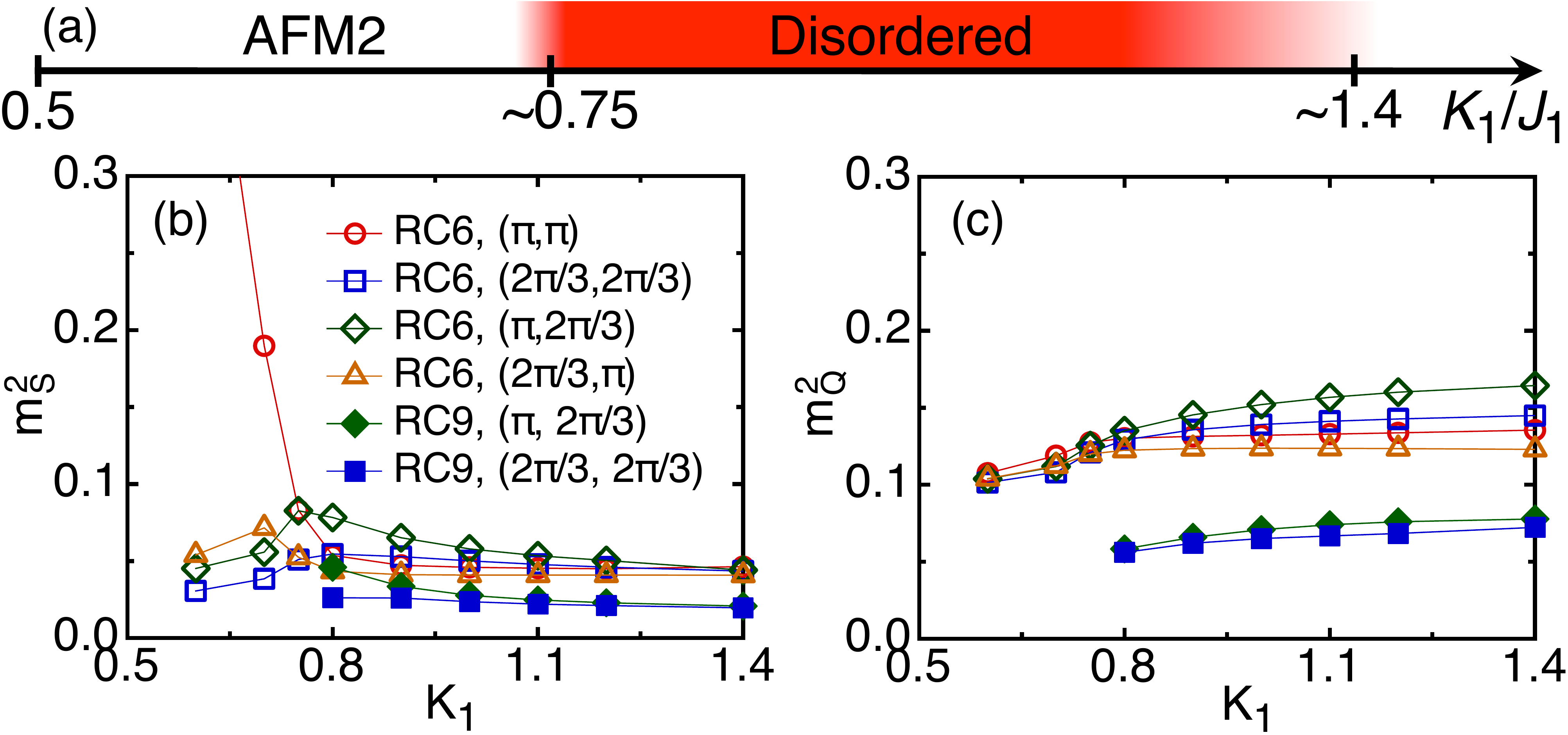}
\caption{(a) Quantum phase diagram of spin-1 bilinear-biquadratic Heisenberg model on the square lattice obtained from our DMRG calculations. The disordered regime around the SU(3) point is indicated in red shading. (b) and (c) are the spin ($m^2_S$) and quadrupolar ($m^2_Q$) order parameters at four dominant momenta as a function of $K_1$, which are obtained from the middle $L_y\times 2L_y$ sites on the long cylinders with $L_y=6,9$.}\label{phase}
\end{figure}

\begin{figure}
\includegraphics[width = 1.0\linewidth]{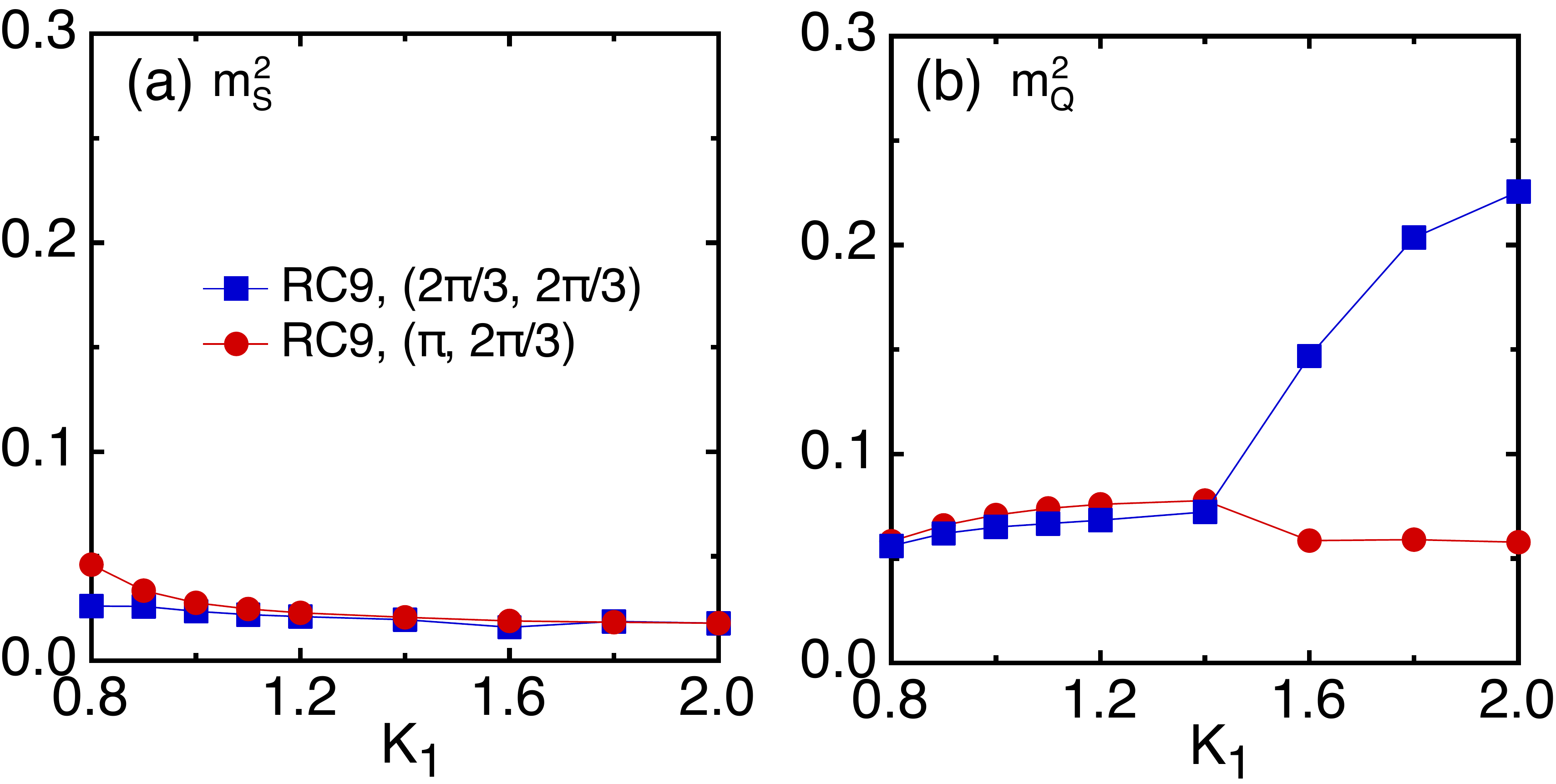}
\caption{Spin ($m^2_S$) and quadrupolar ($m^2_Q$) order parameters at momenta $(\pi,2\pi/3)$ and $(2\pi/3,2\pi/3)$ as a function of $K_1 (K_1 \geq 0.8)$, which are obtained from the middle $L_y\times 2L_y$ sites on the long cylinders with $L_y=9$.}\label{allk}
\end{figure}

Previous studies found an AFM2 phase for $K_1 \lesssim 0.75$ and an AFM3 phase for $0.75 \lesssim K_1 \leq 1.0$~\cite{toth2010,toth2012, bauer2012}. Recent iPEPS simulation also reported a small Haldane phase above $K_1 \simeq 0.75$, between the AFM2 and AFM3 phase~\cite{corboz2017}. We show our phase diagram in Fig.~\ref{phase}(a). We find an AFM2 phase for $K_1 \lesssim 0.75$ and a disordered phase for $0.75 \lesssim K_1 \lesssim 1.4$. We characterize the phases using spin and quadrupolar order parameters $m^{2}_{S}({\bf q}) = \frac{1}{N_s^2}\sum_{i,j} \langle {\bf S}_{i}\cdot {\bf S}_{j} \rangle e^{i{\bf q}\cdot({\bf r}_i-{\bf r}_j)}$ and $m^{2}_{Q}({\bf q}) = \frac{1}{N_s^2}\sum_{i,j} \langle {\bf Q}_{i}\cdot {\bf Q}_{j} \rangle e^{i{\bf q}\cdot({\bf r}_i - {\bf r}_j)}$ (${\bf Q}_{i}$ is the quadrupolar operator~\cite{blume1969}), where the sites $i, j$ are chosen over the middle $N_s = L_y \times 2L_y$ sites in order to avoid edge effects~\cite{gong2014square}. In Figs.~\ref{phase}(b-c), we show the peaks of both orders at four dominant momenta as a function of $K_1$. At $K_1 \simeq 0.75$, the AFM2 order $m_{S}^{2}(\pi,\pi)$ drops sharply, suggesting a first-order transition with vanished AFM2 order, which agrees with previous result~\cite{toth2012}. In Fig.~\ref{allk}, we demonstrate magnetic and quadrupolar orders for larger $K_1$. For $K_1 \gtrsim 1.4$, the AFQ order $m^2_Q$ at ${\bf q} = (2\pi/3,2\pi/3)$ starts to enhance and the AFM order is still vanished, which suggests an emergent AFQ3 order and agrees with previous result~\cite{toth2012}. In this work, we will focus on the most intriguing regime for the intermediate $0.75 \lesssim K_1 \lesssim 1.4$.

\begin{figure}[t]
\includegraphics[width = 1.0\linewidth]{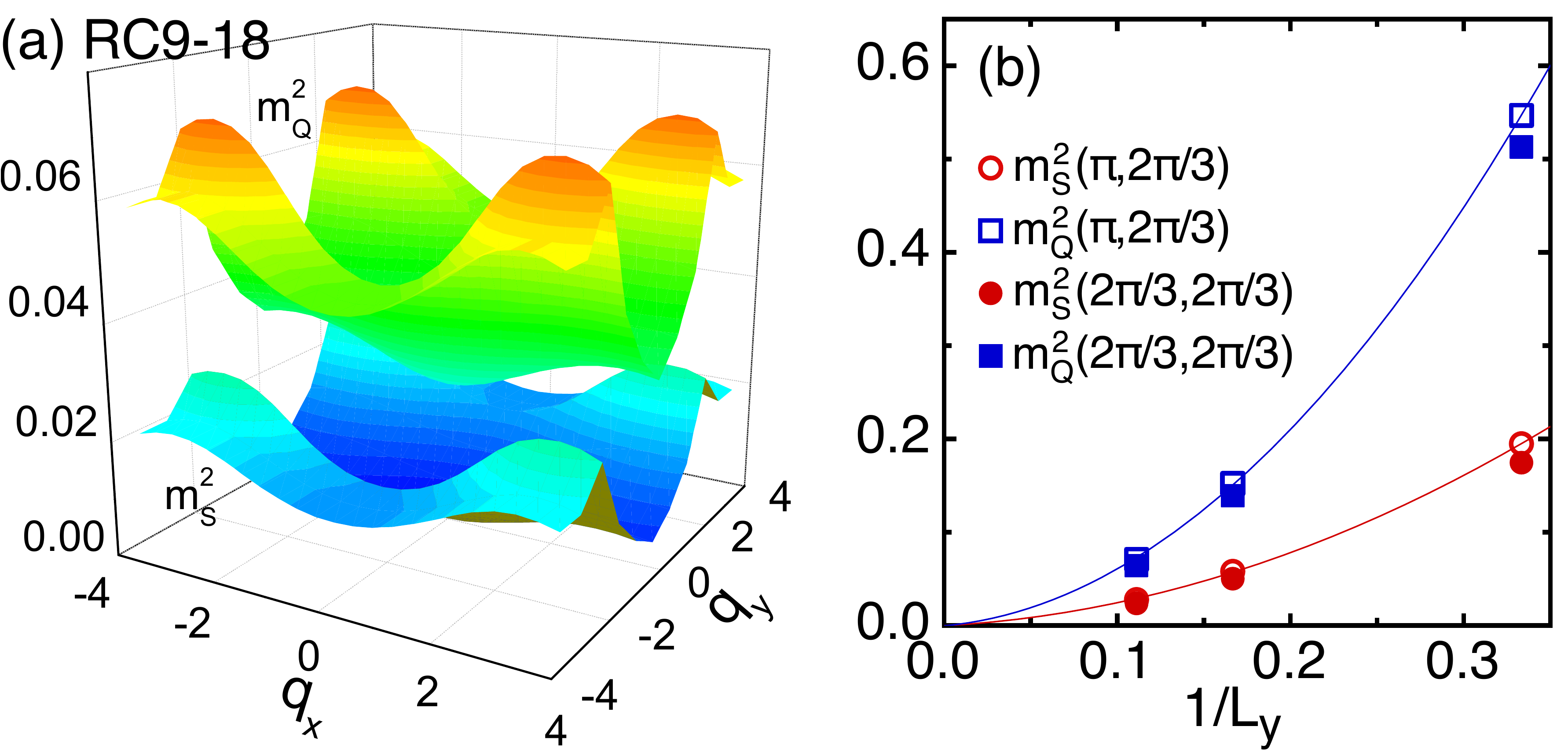}
\caption{Spin and quadrupolar orders for the SU(3) model. (a) Momentum dependence of spin ($m^2_S$) and quadrupolar ($m^2_Q$) orders on RC9, which are obtained from the middle $9 \times 18$ sites. (b) Finite-size scaling of the ordering peaks at ${\bf q} = (\pi,2\pi/3)$ and $(2\pi/3,2\pi/3)$ with $L_y = 3, 6, 9$. The lines are guides to the eye.
}\label{su3}
\end{figure}

In this intermediate phase, we find that the choice of cylinder geometry is important in DMRG calculation: (1) On the RC cylinder, circumference should be the multiple of $3$, i.e. $L_y = 3n$ ($n$ is an integer), to accommodate the dominant structure factor at ${\bf q} = (\pi, 2\pi/3)$ as shown in Figs.~\ref{phase}-\ref{su3}. (2) We note that the shifted cylinders used in Ref.~\onlinecite{bauer2012}, which connect the top site of the $i$-th column to the bottom site of the $(i+1)$-th column, favor the momentum $2\pi/3$ but frustrate the momentum $\pi$. We compare the bulk energy of the SU(3) model on the RC and shifted cylinders, showing that the RC cylinders with $L_y = 3n$ have the lowest energy. For example, the energy per site on the RC9 is around $-0.652$, which is close but $\sim 5\%$ lower than that of the AFM3 state on the shifted RC8 cylinder ($-0.625$ from Ref.~\onlinecite{bauer2012}). Therefore, we focus on the unrestricted RC cylinders with $L_y = 3, 6, 9$, which can harbor the momenta at both $(\pi, 2\pi/3)$ and $(2\pi/3, 2\pi/3)$. We also compare our energy of the SU(3) model with that in the iPEPS simulation. For the ground-state energy of the SU(3) model, our DMRG result on the largest size $L_y = 9$ is $-0.652$, and the iPEPS energy by keeping the most bond dimension $D = 20$ is $-0.6504$~\cite{energy}, showing the good performance of our DMRG results.

\subsection{Identification of the disordered nematic phase} 

For $0.75 \lesssim K_1 \lesssim 1.4$, we study the finite-size scaling of order parameters on the RC3, RC6, and RC9 cylinders. We find that both spin and quadrupolar orders decay fast and properly extrapolate to vanish in the thermodynamic limit, as shown in Fig.~\ref{su3}(b) for the SU(3) point (see Fig.~\ref{k09k11} in Appendix~\ref{app2} for other $K_1$). Thus, we find no formation of either a spin or quadrupolar order in the intermediate phase. 

\begin{figure}
\includegraphics[width = 1.0\linewidth]{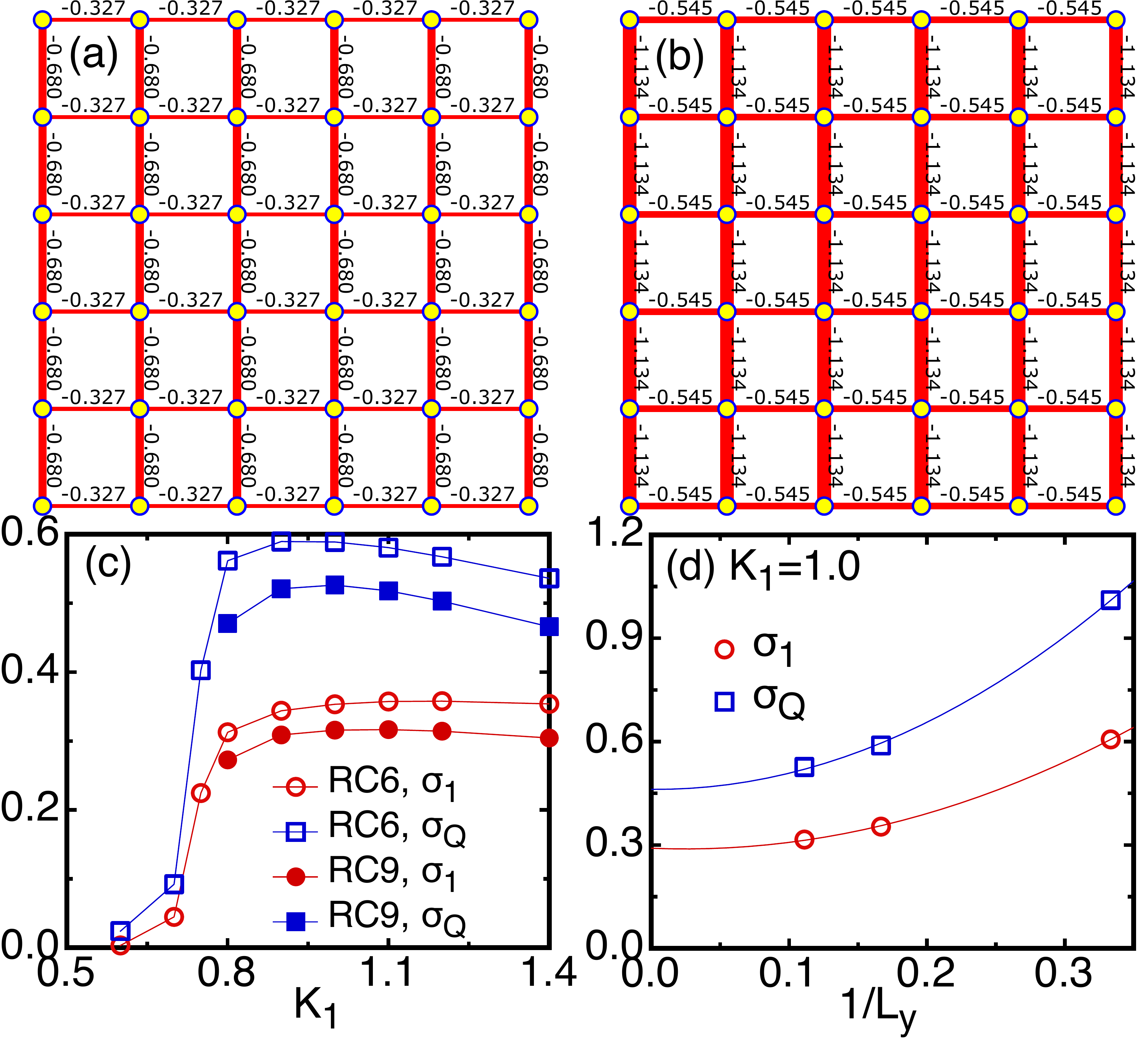}
\caption{Lattice rotational symmetry breaking in the disordered phase. (a) and (b) are the nearest-neighbor $\langle {\bf S}_i \cdot {\bf S}_j \rangle$ and $\langle {\bf Q}_i \cdot {\bf Q}_j \rangle$ for the SU(3) model in the bulk of RC6. (c) Nematic order parameters $\sigma_1$ and $\sigma_Q$ as a function of $K_1$ on RC6 and RC9. (d) Finite-size scaling of $\sigma_1$ and $\sigma_Q$ for the SU(3) model with $L_y = 3, 6, 9$.}
\label{sigma}
\end{figure}

Next, we examine the lattice translation symmetry by calculating the nearest-neighbor dipolar bond energy $\langle {\bf S}_{i}\cdot {\bf S}_{j}\rangle$ and quadrupolar bond energy $\langle {\bf Q}_{i}\cdot {\bf Q}_{j}\rangle$. We find that the bond energy differences along the $x$ direction all decay quite fast from the edge to the bulk, leading to the translationally uniform bond energy in the bulk as shown in Figs.~\ref{sigma}(a-b) on RC6 (the same for RC3 and RC9). The uniform bond energy indicates the preserved translational symmetry, precluding the possibility of a VBC order. Note that the absence of the VBC order on the square lattice is different from the kagome and honeycomb SU(3) models, in which the ground states have been identified as breaking lattice symmetries~\cite{corboz2012, zhao2012, corboz2013}.

We now turn to analyze the nematic order. In Figs.~\ref{sigma}(a-b), one can see a strong anisotropy between horizontal and vertical bond energies. We define the nematic order parameters $\sigma_1$ and $\sigma_Q$ as $\sigma_1 \equiv \frac{1}{N_m}\sum_i[\langle {\bf S}_{i}\cdot {\bf S}_{i+\hat{x}}\rangle - \langle {\bf S}_{i}\cdot {\bf S}_{i+\hat{y}}\rangle]$~\cite{Chandra1990} and $\sigma_Q \equiv \frac{1}{N_m}\sum_i[\langle {\bf Q}_{i}\cdot {\bf Q}_{i+\hat{x}}\rangle - \langle {\bf Q}_{i}\cdot {\bf Q}_{i+\hat{y}}\rangle]$~\cite{rong2015} ($\hat{x}$ and $\hat{y}$ denote the unit vectors along the two directions, $N_m$ is the number of sites of the two columns in the middle of cylinder). The results of $\sigma_1$ and $\sigma_Q$ versus $K_1$ are illustrated in Fig.~\ref{sigma}(c). Upon the transition at $K_1 \simeq 0.75$, both $\sigma_1$ and $\sigma_Q$ grow dramatically. Since finite-size scaling of nematic order has been shown as an efficient method to identify the lattice $C_4$ symmetry breaking for different quantum phases in DMRG calculation~\cite{hufese}, we extrapolate our data as shown in Fig.~\ref{sigma}(d), which clearly obtain nonzero $\sigma_1$ and $\sigma_Q$ in the thermodynamic limit. Our results indicate that the $C_4$ symmetry breaking occurs spontaneously and is not the result of the cylindrical lattice geometry.

In order to further characterize the nature of the disordered phase, we calculate the spin gap $\Delta_{T}$ with $\Delta S = 1, 2$ ($S$ is the total spin quantum number) as well as the spin-singlet gap in the bulk of a long cylinder by sweeping the excited states in the bulk~\cite{yan2011}. The two spin gaps $\Delta S = 1,2$ have the same value at the SU(3) point, as anticipated from the SU(3) symmetry (see Fig.~\ref{gap6} in Appendix~\ref{app2}). Limited by the system size on the RC3 and RC6, we do not perform finite-size scaling for the gaps. However, we remark the fast drop of the gaps from RC3 to RC6, with the singlet gap reducing from $1.853$ to $0.285$ (more than $80\%$) and the triplet gap from $2.028$ to $0.774$ (more than $60\%$), which suggests either tiny or vanishing gaps in the thermodynamic limit.

\section{Discussion on the nature of the disordered phase}
\label{sec:discussion}
 
According to the Lieb-Schultz-Mattis-Hastings theorem~\cite{lieb1961,Hastings2004}, this spin-$1$ disordered phase can be either a gapped quantum paramagnet with a unique ground state, a gapped topological spin liquid, or a gapless spin liquid. We carefully study these possibilities here.  
 
\subsection{Space anisotropic model} 
 
\begin{figure}
\includegraphics[width = \linewidth]{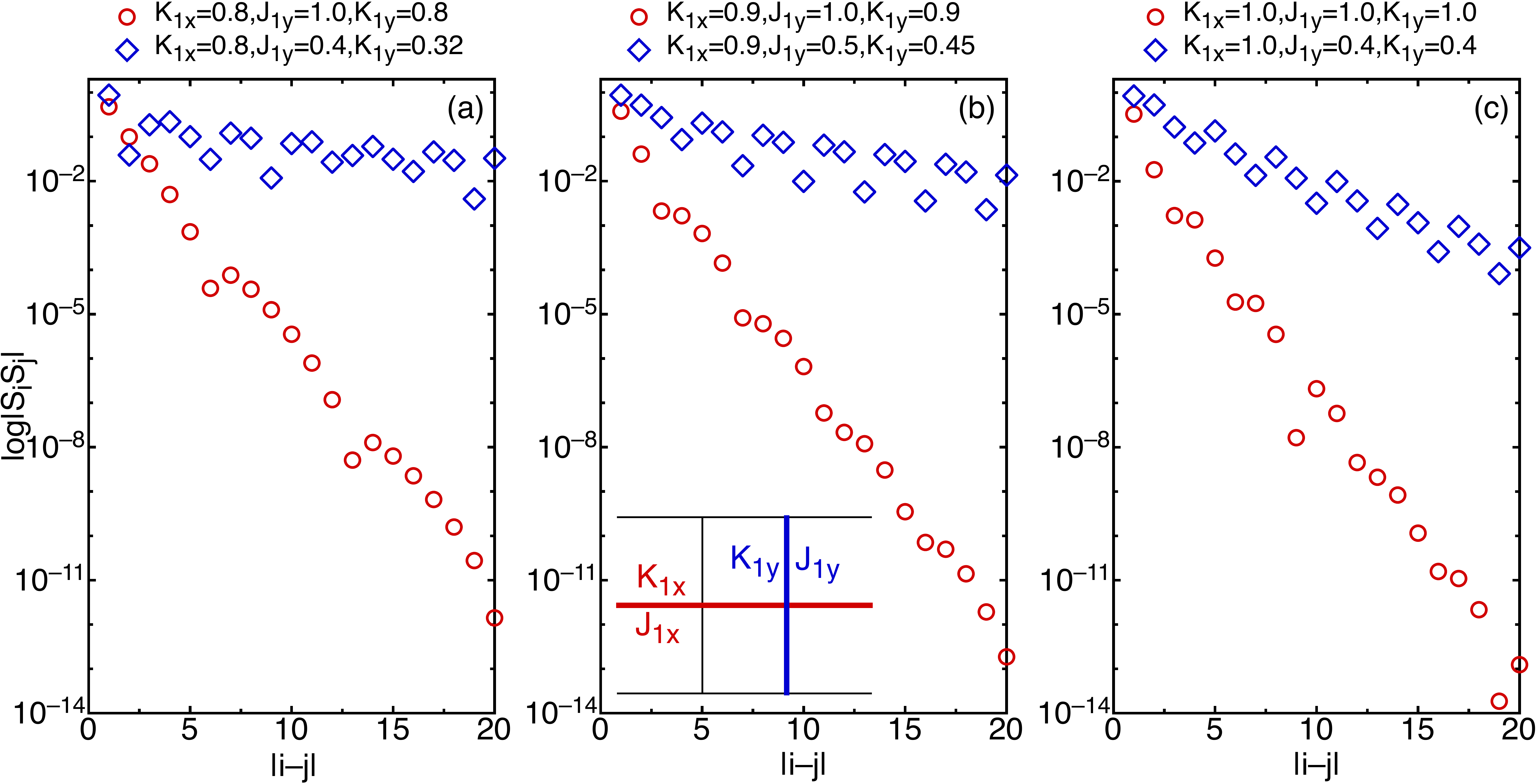}
\caption{Log-linear plots of spin-spin correlations as a function of the distance on the RC6 cylinder for coupling chains with different intrachain and interchain bilinear-biquadratic interactions: $J_{1x}=1.0$, (a) $K_{1x}=0.8$ and $J_{1y}/J_{1x}=K_{1y}/K_{1x}=0.4$, (b) $K_{1x}=0.9$ and $J_{1y}/J_{1x}=K_{1y}/K_{1x}=0.5$, (c) $K_{1x}=1.0$ and $J_{1y}/J_{1x}=K_{1y}/K_{1x}=0.4$.}\label{Fig:ani}
\end{figure}
  
First of all, we study the space anisotropic model with $K_{1x}/J_{1x} = K_{1y}/J_{1y}$ ($J_{1x}, K_{1x}$ and $J_{1y}, K_{1y}$ are the couplings along the $x$ and $y$ directions), as shown in the inset of Fig.~\ref{Fig:ani}(b).
We set $J_{1x}=1.0$ as energy scale and for a given $K_{1x}$, we vary $J_{1y}$ and $K_{1y}$.
With equal interchain and intrachain couplings, the system is the studied isotropic model. 
Without interchain couplings, the system reduces to decoupled spin-$1$ chains which is in the gapped Haldane phase for $K_{1x}/J_{1x} < 1$~\cite{aklt1987}. 

We thus focus on the regime $0.75 <K_{1x} / J_{1x} \leq 1$, and increase the interchain couplings $J_{1y}, K_{1y}$. 
Near the isotropic model, there is no long-range magnetic order (red symbols in Fig.~\ref{Fig:ani}), as is expected in the spin liquid phase. 
In the intermediate interchain couplings, we find that the spin correlations clearly show an emergent AFM order with slowly decaying correlations as shown by the blue symbols in Fig.~\ref{Fig:ani}. 
Therefore, we find that the disordered phase in the isotropic model is not smoothly connected to the Haldane phase, strongly suggesting that the disordered phase we found is not an adiabatic extension of the 1d-limit of weakly coupled Haldane chains. 
We note that a Haldane phase has been proposed in a very small region above the transition at $K_1/J_1 \simeq 0.75$ in Ref.~\onlinecite{corboz2017}. 
It may be possible that a Haldane-like phase exists in a narrow parameter regime between the N\'eel magnetic order phase and the nematic spin liquid phase found here, although we have not found evidence thereof in our DMRG results. 

\subsection{Possible a gapless spin liquid with emergent Fermi points}

\begin{figure}
\includegraphics[width = 0.9\linewidth]{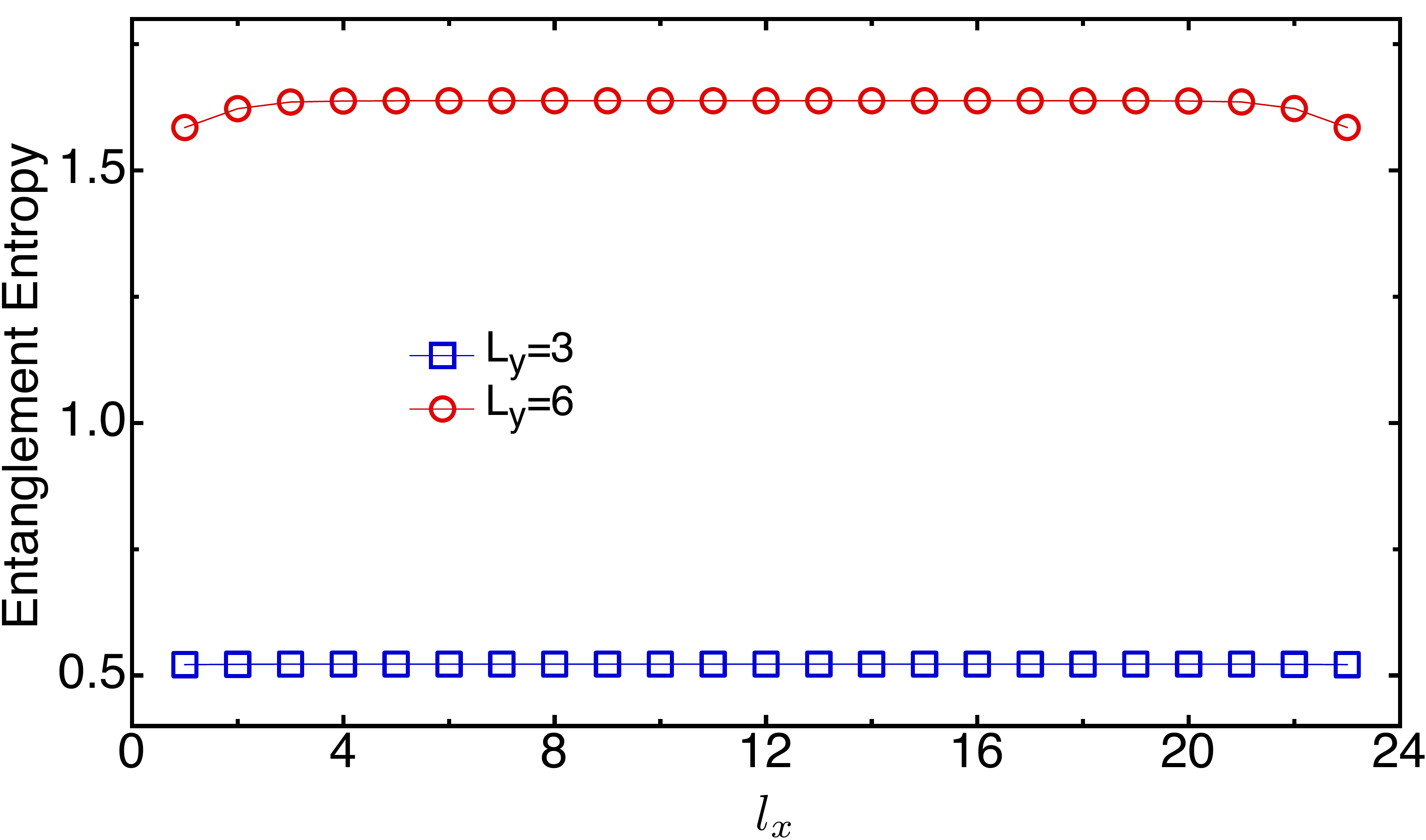}
\caption{von Neumann entanglement entropy versus the subsystem length $l_x$ for the SU(3) model on the RC3-24 and RC6-24 cylinders.}\label{Fig:entropysu3}
\end{figure}

Furthermore, we take advantage of the established methodology in gapped spin-$1/2$ spin liquids, where one can test the spinon topological sector by removing or adding a spin-$1/2$ on each open edge of cylinder~\cite{yan2011}. We follow this prescription by adding or removing a spin-$1$ site on each open edge of the RC3 and RC6 cylinders to detect possible spinon sector in the disordered phase. However, we do not find near-degenerate spinon sector (see Fig.~\ref{Fig:bulk6} in Appendix~\ref{app2}), which does not support a gapped spin liquid.

The numerical results and the discussions above motivate us to consider a gapless phase. If the gapless spin liquid possesses a Fermi surface formed by the emergent or strongly renormalized degrees of freedom~\cite{Lee2006}, $m_S({\bf q})$ and $m_Q({\bf q})$ should show singular peaks at ${\bf q} = \bm{k}^{\hat{n}}_{FR} - \bm{k}^{\hat{n}}_{FL}$, where $\bm{k}_{FR/L}$ are the locations of the Right/Left patches of the Fermi surface perpendicular to an observation direction $\hat{n}$. 
On the $L_y$-leg ladder with periodic boundary conditions along the $\hat{y}$ direction, there are $L_y$ lines cutting through the momentum space. 
As the \textit{fingerprints} of the ladder descendant of the 2D Fermi surface, we expect to see the singular behavior at $\bm{q}^i = \bm{k}^i_{FR} - \bm{k}^i_{FL}$, where $i = 1,2,\cdots,~L_y$ corresponding to the $i$-th line and $\bm{k}^i_{FR/L}$ represent the Right/Left Fermi points of the Fermi surface intersecting the line. 
However, the DMRG results in Fig.~\ref{su3} show that on a $9$-leg cylinder $m^2_S(\bm{q})$ and $m^2_Q(\bm{q})$ only possess rather broad (instead of singular) peaks, indicating that there is no emergent Fermi surface.

For an alternate route to study the gapless nature, we calculate the von Neumann entanglement entropy by dividing the system to two subsystems and computing the eigenvalues $\lambda_i$ of the reduced density matrix, which leads to the entropy $S = -\sum_i \lambda_i \ln \lambda_i$.
We find that the entropy as a function of the subsystem size $l_x$ is invariant, which does not show any logarithmic correction that is expected for a Fermi surface (see Fig.~\ref{Fig:entropysu3}). 
These results suggest that a spinon Fermi surface is absent~\cite{sheng2009} and, instead, emergent Fermi points are more likely. 
For such a state on finite size system, the Fermi points of spinon may not be touched and the entropy would be flat with growing subsystem length. 
Due to the limitation of system size, this issue certainly requires further scrutiny such as variational Monte Carlo calculation with Gutzwiller projected parton constructions.

In the frame of gapless spin liquid, the nematic nature can be understood based on the low-energy gapless fluctuations. For low energy description, the spin operator can be expressed as $S_j \sim e^{i \bm{q} \cdot \bm{r}_j} \mathcal{S}_{\bm{q}} + H.c.$, where the vector $\bm{q}$ is the wavevector associated with the gapless excitations. The bond energy operator can be expressed as $\mathcal{B}_{\bm{\delta}} = S_j \cdot S_{j + \bm{\delta}} \sim e^{ i \bm{q} \cdot \bm{\delta}} \mathcal{S}_{\bm{q}} \mathcal{S}_{- \bm{q}} + H.c.$, where $\bm{\delta}$ is the unit vector corresponding to $(1,0)$ or $(0,1)$. The difference between the bond energies along $\hat{x}$ and $\hat{y}$ directions is $\Delta\mathcal{B} \equiv \mathcal{B}_x - \mathcal{B}_y = \left( e^{i q_x} - e^{i q_y} \right) \mathcal{S}_{\bm{q}} \cdot \mathcal{S}_{-\bm{q}}  + H.c.$. For a vector $\bm{q}$ with unequal magnitudes of $x$ and $y$ components, such as $\bm{q} = (\pi,2\pi/3)$ in this model, $\left\langle \Delta \mathcal{B} \right\rangle \not = 0$ leads to a nonzero nematic order. The bond-energy associated with the quadrupolar moments shows a similar effect.

We stress that this mechanism for the nematic order operates not only for static order such as the AFQ23, but also for fluctuations of such order. Our DMRG calculations have indeed found the $(\pi,2\pi/3)$ fluctuations at $L_y$-leg ladder near the SU(3) point, shown in Figs.~\ref{phase}-\ref{su3}. The precise origin of this wavevector $(\pi,2\pi/3)/(2\pi/3,\pi)$ remains to be understood, but it is already hinted at by the local stability of such an order within the semiclassical analysis described earlier. In addition, the tendency towards dominant fluctuations of such a wavevector arises from a coupled-chain picture of the gapless spin liquid: $2\pi/3$ is twice of the parton ``Fermi-wavevector" for an SU(3)-chain, which is associated with the $1/3$-filling in terms of the partons, and $\pi$ reflects the nearest-neighbor coupling being the interchain coupling that is relevant in the RG sense (see Appendix~\ref{app3}). 

\section{Discovering the AFM3 phase}
\label{sec:afm3}

\begin{figure}
\includegraphics[width = \linewidth]{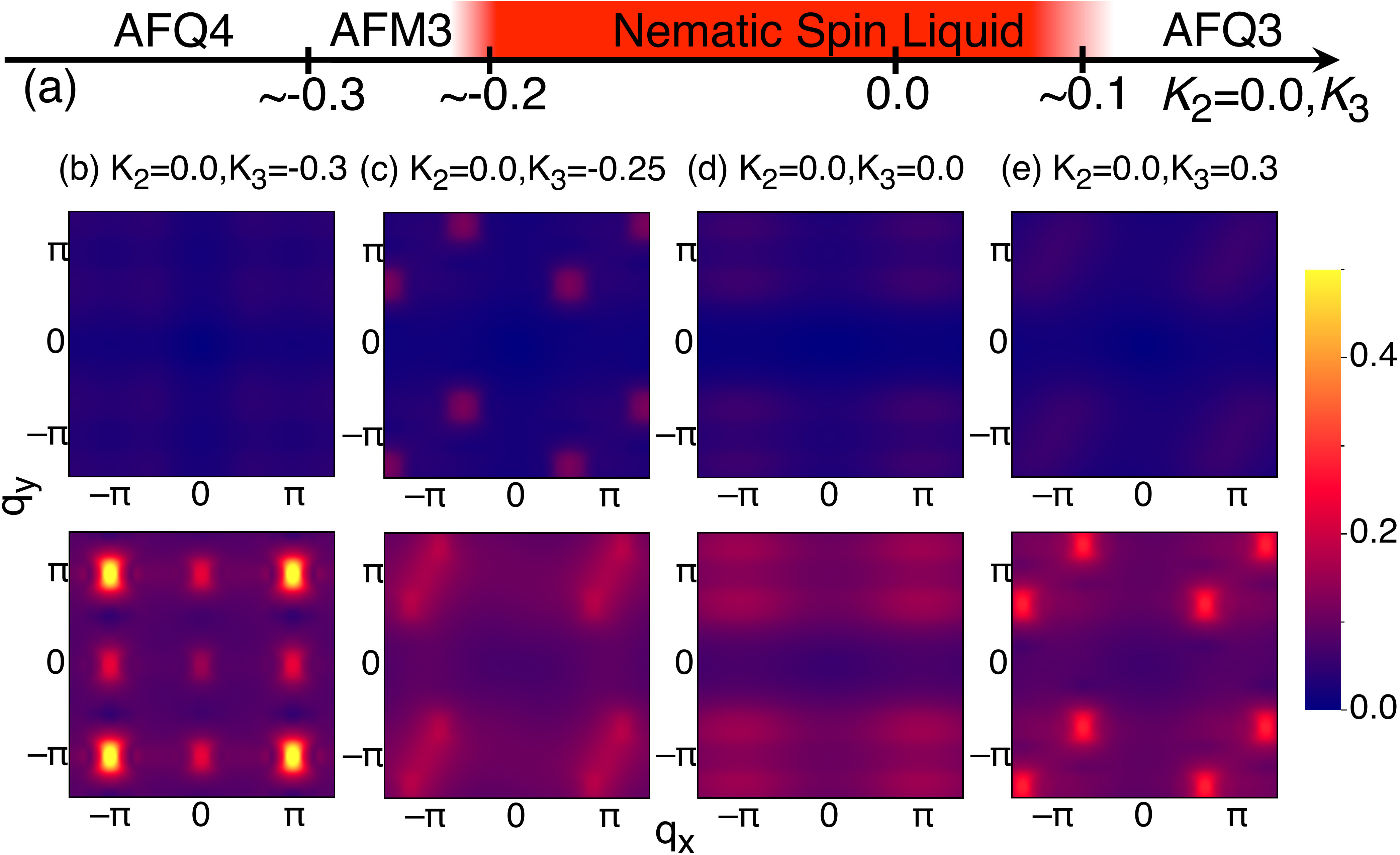}
\caption{(a) Quantum phase diagram of the spin-$1$ bilinear-biquadratic Heisenberg model on the square lattice with changing $K_3$ and $K_2=0$. The regime of the nematic spin liquid around the SU(3) point is indicated in red shading. (b)-(e) are the spin ($m^2_S$) and quadrupolar ($m^2_Q$) structure factors for different $K_3$ with $K_2=0.0$, which are obtained from the middle $6\times 12$ sites on the long RC6 cylinder. The upper and lower figures are for $m^2_S$ and $m^2_Q$, respectively.} \label{pdk3}
\end{figure}

In previous numerical calculation of the SU(3) model~\cite{bauer2012}, the ground state has been found as an AFM3 magnetic order state. In our study, we identify the ground state of the system as a spin liquid state. However, on the TC cylinder (see Appendix~\ref{app2}) and the shifted RC cylinder that was used in Ref.~\onlinecite{bauer2012}, the system indeed favors an AFM3 order. Although the spin liquid state has the lower energy than the AFM3 state, the AFM3 order seems to have a competitive energy. Therefore, it is reasonable to expect that the AFM3 order could be found near the spin liquid phase by tuning some coupling parameters. We have numerically tested the SU(3) model with different additional interactions, and we found the AFM3 order by considering the third-neighbor biquadratic coupling $K_3$ ($K_2 = 0$), as shown in the quantum phase diagram Fig.~\ref{pdk3}. The structure factors of Figs.~\ref{pdk3}(b)-(e) clearly show at least three quantum ordered phases, including the AFQ4, the AFM3, and AFQ3 phase. In a small parameter region $-0.3 \lesssim K_3 \lesssim -0.2$, one can see the structure factor peaks of the AFM3 order in Fig.~\ref{pdk3}(c).

\section{Summary}
\label{sec:summary}

We have studied the ground state of the spin-$1$ bilinear-biquadratic model on the square lattice using semiclassical flavor-wave theory and extensive DMRG calculation. We find a disordered phase for $0.75 \lesssim K_1/J_1 \lesssim 1.4$ by showing vanished spin and quadrupolar orders, as well as absent translational symmetry breaking. The bond energy texture indicates a spontaneous breaking of $C_4$ lattice rotational symmetry. Based on our results, we suggest this disordered phase as a nematic spin liquid. The spin triplet gap drops rapidly with system size, which strongly suggests gapless excitations. We discuss the origin of the lattice nematic order and the dominant $(\pi, 2\pi/3)$ fluctuations in the framework of gapless spin liquid. By considering additional further-neighbor couplings, we find that this spin liquid phase is stable in a range with different changing couplings. We also find the previously proposed AFM3 phase, which is at the neighbor of the spin liquid phase, in presence of a small third-neighbor biquadratic interaction $K_3 < 0$. Our results lead to a hitherto unexplored mechanism for quantum nematic spin liquid and, more generally, open up a new route towards novel phases of quantum spin systems. For further understanding on this exotic nematic spin liquid phase, variational wavefunction study and renormalization group simulation on larger system size are highly demanded.
\\

\begin{center}
{\bf ACKNOWLEDGMENTS}
\end{center}

We thank Andreas Ludwig, Chao-Ming Jian, Cenke Xu, Donna Sheng, Fr\'ed\'eric Mila, Federico Becca, Kun Yang, Meng Cheng, Matt Foster, Olexei I. Motrunich, Philippe Corboz, Rong Yu, Subir Sachdev, and Tim Hsieh for fruitful discussions. The work was supported in part by the NSF Grant No.\ DMR-1350237 (W.-J.H., H.-H.L. and A.H.N.), the U.S. Department of Energy, Office of Science, Basic Energy Sciences, under Award No. DE-SC0018197 and the Robert A.\ Welch Foundation Grant No.\ C-1411 (W.-J.H., H.-H.L., H. H., and Q.S.), a Smalley Postdoctoral Fellowship of the Rice Center for Quantum Materials (H.-H. L.), Cottrell Scholar Award from the Research Corporation for Science Advancement (W.-J.H. and A.H.N.), the NSFC Grant No.\ 11874078, 11834014 (S.-S.G.), the Fundamental Research Funds for the Central Universities (S.-S.G.), the National High Magnetic Field Laboratory through the NSF Grant No.\ DMR-1157490 and the State of Florida (S.-S.G.), and the Robert A.\ Welch Foundation Grant No.\ C-1818 (A.H.N.). The majority of the computational calculations have been performed on the Shared University Grid at Rice funded by NSF under Grant EIA-0216467, a partnership between Rice University, Sun Microsystems, and Sigma Solutions, Inc., the Big-Data Private-Cloud Research Cyberinfrastructure MRI-award funded by NSF under Grant No. CNS-1338099 and by Rice University, the Extreme Science and Engineering Discovery Environment (XSEDE) by NSF under Grant No.\ DMR160057. 

\appendix

\section{Flavor wave theory calculations}
\label{app1}

\begin{figure}[t]
\includegraphics[width = 3 in]{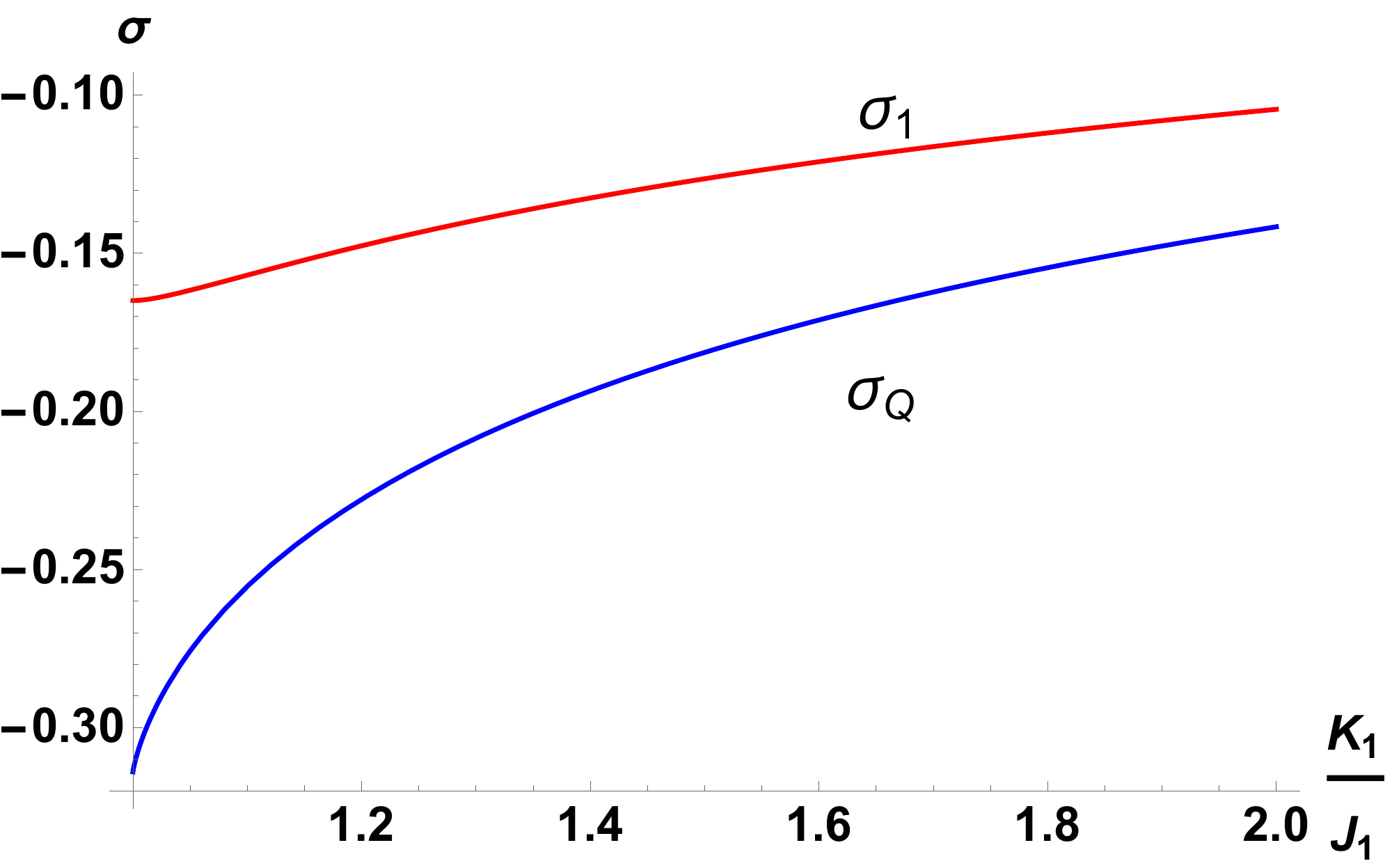}
\caption{Nematic orders $\sigma_1$ and $\sigma_Q$ in AFQ23 within the flavor-wave theory calculations. We can see that the nematic orders are \textit{finite} indicating the broken C$_4$ lattice rotation. }
\label{supp:fig_afqnem}
\end{figure}

Here, we demonstrate the details of the flavor wave theory calculation.
We first give the dispersions associated with the gapless Goldstone modes together with the nematic order parameters in different orders. Two types of nematic order parameters are considered in this work, defined as
\begin{eqnarray}
&& \sigma_1 \equiv \frac{1}{N_s} \sum_i \left[\left\langle \bm{S}_i \cdot \bm{S}_{i + \hat{x}} \right\rangle - \left\langle \bm{S}_i \cdot \bm{S}_{i + \hat{y}}\right\rangle\right],\\
&& \sigma_Q \equiv \frac{1}{N_s} \sum_i \left[\left\langle \bm{Q}_i \cdot \bm{Q}_{i + \hat{x}} \right\rangle - \left\langle \bm{Q}_i \cdot \bm{Q}_{i + \hat{y}}\right\rangle\right],
\end{eqnarray}
where $N_s$ represents the number of sites, $i$ is the site labelings, and the quadrupolar bond energy can be re-expressed in terms of the spin bond energy using the identity $\bm{Q}_i \cdot \bm{Q}_j = 2 (\bm{S}_i \cdot \bm{S}_j )^2 + \bm{S}_i \cdot \bm{S}_j - 2 \bm{S}_i^2 \bm{S}_j^2/3$.
\begin{enumerate}
\item[(1)] FM: \\
 After the standard procedure, we find that the boson Hamiltonian is
 \begin{eqnarray}
 H = \sum_{\bm{k},\alpha = 0,\bar{1}} \varepsilon_{\bm{k}\in BZ,\alpha} 
\left( b^\dagger_{\bm{k},\alpha}b_{\bm{k},\alpha} + \frac{1}{2}\right) + 8 N_s J_1,
 \end{eqnarray}
 where BZ stands for the Brillouin zone, and $N_s$ is the total number of sites, and the dispersions are
 \begin{eqnarray}
 && \varepsilon_{\bm{k},0} = 2 J_1 \left[ \cos\left(k_x\right) + \cos\left( k_y\right) -2 \right],\\
 && \varepsilon_{\bm{k},\bar{1}} = 2K_1 \left[ \cos\left( k_x \right) + \cos\left(k_y \right)\right] + 4 \left( K_1 - 2 J_1\right).
\end{eqnarray}
 The nematic orders in FM are obviously zero ($\sigma_1 = \sigma_Q = 0$) since there is no rational symmetry breaking in FM.
 \item[(2)] AFM2 (N\'eel AFM): \\
 The boson Hamiltonian after performing Fourier transform is
 \begin{eqnarray}
\nonumber H=\sum_{\bm{k},\alpha\in{0,\bar{1}}} \varepsilon_{\bm{k},\alpha} \left( a_{\bm{k},\alpha}^\dag a_{\bm{k},\alpha}+\frac{1}{2}\right)+N_s\left(\frac{16K}{3}-8J\right), \\
\end{eqnarray}
where $N_s$ is the number of sites and the dispersions associated with $a_{\bm{k}, 0}$ and $a_{\bm{k},\bar{1}}$ are
\begin{eqnarray}
\varepsilon_{\bm{k}, 0}=\sqrt{A_{\bm{k},0}^2 - B_{\bm{k},0}^2}, &~~\varepsilon_{\bm{k},\bar{1}}=\sqrt{A_{\bm{k},\bar{1}}^2 - B_{\bm{k},\bar{1}}^2},
\end{eqnarray}
where we define
\begin{eqnarray}
&& A_{\bm{k}, 0}=4(J-K), \\
&& A_{\bm{k},\bar{1}}=4(2J-K), \\
&& B_{\bm{k}, 0}=2(J-K)\left(\cos(k_x)+\cos(k_y)\right),\\
&& B_{\bm{k},\bar{1}}=2K\left(\cos(k_x)+\cos(k_y)\right).
\end{eqnarray}
Since there is no rotational symmetry breaking in AFM2, the nematic orders are zero ($\sigma_1 = \sigma_Q = 0$).
\item[(3)] FQ: \\
 The boson Hamiltonian after performing Fourier transform is
 \begin{eqnarray}
H= \sum_{\bm{k},\alpha\in{0,\bar{1}}} \varepsilon_{\bm{k},\alpha} \left( a_{\bm{k},\alpha}^\dag a_{\bm{k},\alpha}+\frac{1}{2}\right) + 8 N_s K_1,~~~
\end{eqnarray}
where $N_s$ is again the number of sites and we find that the dispersions associated with $a_{\bm{k}, 0}$ and $a_{\bm{k},\bar{1}}$ are the same $\varepsilon_{\bm{k},0} = \varepsilon_{\bm{k},\bar{1}} \equiv \epsilon_{\bm{k}}^{FQ}$, with
\begin{eqnarray}
\nonumber  \varepsilon^{FQ}_{\bm{k}}&=& 2 \Bigg{(} \left[ \cos\left(k_x\right) + \cos \left(k_y\right) - 2K_1 \right]^2 - \\
 && ~~~ -\left[ \left( J_1 - K_1 \right) \left( \cos\left( k_x \right) + \cos \left( k_y \right)\right) \right]^2 \bigg{)}^{\frac{1}{2}}.
\end{eqnarray}
Since FQ does not break rotational symmetry, the nematic order are zero.
\item[(4)] AFQ3: \\
The bosonic Hamiltonian at the SU(3) time-reversal-invariant basis can be concise expressed as
\begin{eqnarray}
H = \sum_{\bm{k}\in BZ,\alpha = y,z} \varepsilon_{\bm{k}, \alpha} \left( a^\dagger_{\bm{k},\alpha} a_{\bm{k},\alpha} + \frac{1}{2}\right),
\end{eqnarray}
with
\begin{eqnarray}
 \varepsilon_{\bm{k},\alpha = y, z} = \sqrt{A^2_{\bm{k},\alpha} - B^2_{\bm{k},\alpha}},
\end{eqnarray}
where we define
\begin{eqnarray}
&& A_{\bm{k},y} = 2K_1 + \left( K_1 - J_1\right)\sqrt{2\left( 1 + \cos\left( k_x + k_y \right)\right)}, \nonumber \\
&&  B_{\bm{k},y} = J_1 \sqrt{2\left( 1 + \cos\left(k_x + k_y \right)\right)},\nonumber \\
&& A_{\bm{k},z} =2K_1 - \left( K_1 - J_1\right)\sqrt{2\left( 1 + \cos\left(k_x + k_y \right)\right)}, \nonumber \\
&& B_{\bm{k},z} = J_1 \sqrt{2\left( 1 + \cos\left(k_x + k_y \right)\right)}.
\end{eqnarray}
We note that the summation in momentum $\bm{k}$ is over the whole BZ for two bands. We can see that along the line with $k_x + k_y = 0$, the $z$-bosons have gapless lines along $\varepsilon_{\bm{k},z}|_{k_x + k_y = 0} = 0$. Due to this gapless line in the whole regime at $K_1/J_1 \geq 1$, the quantum corrections to the ordered moment extracted within the flavor-wave theory are divergent, which hints that the flavor-wave analysis may break down in this regime in this model. At the SU(3) point, the divergence of the quantum correction to the order moment within the flavor-wave picture was also reported previously \cite{bauer2012}. The nematic orders vanish, $\sigma_1=\sigma_Q=0$.
\item[(5)] AFQ23:\\
In the AFQ23, a unit cell contains $2\times 3$ sublattices as shown in Fig.~\ref{fig_order}(b). In our calculations, we choose the $2$ sublattices in the first column to be a primitive cluster (sublattices $A_1$ and $B_2$) and perform local rotations on different sites in the unit cell to align with the states on the first column. Since the AFQ23 has never been considered previously, we provide more details of the calculations here. Focusing on the sublattices in the primitive clusters denoted as $1$ for the original $A_1$ and $2$ for the original $B_2$, we assume that at $1$ only $a_x$-bosons condense and at $2$ only $a_y$-bosons condense, and perform local rotations for $a_{\bm{r},\alpha=x,y,z}(\mu=1,2)$ to transform $a_{\bm{r},\alpha}(\mu) \rightarrow b_{\bm{r},\alpha}(\mu)$, and in the $b$-boson basis, only $b_x$ condense at each site. Explicitly, the local rotations give 
\begin{eqnarray}
&& \Bigg\{
\begin{array}{c}
a_{\bm{r},x} (1)  \\
a_{\bm{r},y}(1) \\
a_{\bm{r},z}(1) 
\end{array} = \Bigg\{ \begin{array}{c}
b_{\bm{r},x}(1) \\
b_{\bm{r},y}(1)\\
b_{\bm{r},z}(1)
\end{array}; \\
&& \Bigg\{
\begin{array}{c}
a_{\bm{r},x} (2)  \\
a_{\bm{r},y}(2) \\
a_{\bm{r},z}(2) 
\end{array} = \Bigg\{ 
\begin{array}{c}
b_{\bm{r},z}(2) \\
b_{\bm{r},x}(2)\\
b_{\bm{r},y}(2)
\end{array} 
\end{eqnarray}
At Fourier space, we find that the Hamiltonian in terms of the bosons is
\begin{widetext}
\begin{eqnarray}
\nonumber H =\sum_{{\bf k} \in RBZ}\Bigg\{ && 3K_1 \left[b_{\bm{k},y}^\dag(1)b_{\bm{k},y}(1)+b_{\bm{k},z}^\dag(2)b_{\bm{k},z}(2) \right]+ K_1 \left[ b_{\bm{k},z}^\dag(1)b_{\bm{k},z}(1)+b_{\bm{k},y}^\dag(2)b_{\bm{k},y}(2) \right] +\\
\nonumber &&+(K_1-J_1) \left[ e^{i k_x} \left(b_{\bm{k},y}^\dag(1)b_{\bm{k},z}(1)+b_{\bm{k},y}^\dag(2)b_{\bm{k},z}(2)\right) + H.c.\right] + \\
\nonumber &&+J_1\left[e^{i k_x} \left( b_{\bm{k},y}^\dag(1)b_{-\bm{k},z}^\dag(1)+b_{\bm{k},y}^\dag(2)b_{-\bm{k},z}^\dag(2)\right) + H.c.\right] + \\
&&+(K_1-J_1)\bigg[(1+e^{i k_y})b_{\bm{k},y}^\dag(1)b_{\bm{k},z}(2)+ H.c.\bigg]+J_1\bigg[(1+e^{i k_y})b_{\bm{k},y}^\dag(1)b_{-\bm{k},z}^\dag(2)+ H.c.\bigg]+4K_1\Bigg\}
\end{eqnarray}
\end{widetext}
Where $RBZ$ means the reduced Brillouin zone, which will be suppressed below for clarity. For diagnosing the Hamiltonian, we first simplify the Hamiltonian with new boson fields, defined as
\begin{eqnarray}
&&b_{\bm{k}, y}^\dag(1)=\frac{e^{-i k_y/2}}{\sqrt{2}}(c_{\bm{k}}^\dag(1) - c_{\bm{k}}^\dag(2))\\
&&b_{\bm{k},z}^\dag(2)=\frac{1}{\sqrt{2}}(c_{\bm{k}}^\dag(1)+c_{\bm{k}}^\dag(2))\\
&&b_{\bm{k},z}^\dag(1)=\frac{e^{i(k_x - k_y/2)}}{\sqrt{2}}(d_{\bm{k}}^\dag(1)-d_{\bm{k}}^\dag(2))\\
&&b_{\bm{k},y}^\dag(2)=\frac{e^{-i k_x}}{\sqrt{2}}(d_{\bm{k}}^\dag(1)+d_{\bm{k}}^\dag(2)),
\end{eqnarray}
followed by a generalized Bogoliubov transformation
\begin{eqnarray}
&&c_{\bm{k}}(1)=u_{\bm{k},1} \gamma_{\bm{k},1}+u_{2,p} \gamma_{\bm{k},2} + v_{\bm{k},1} \gamma_{-\bm{k},1}^\dag +v_{\bm{k},2} \gamma_{-\bm{k},2}, \nonumber \\
&&d_{\bm{k}}(1)=u_{\bm{k},3} \gamma_{\bm{k},1}+u_{\bm{k},4} \gamma_{\bm{k},2} +v_{\bm{k},3} \gamma_{-\bm{k},1}^\dag +v_{\bm{k},4} \gamma_{-\bm{k},2}, \nonumber \\
&&c_{\bm{k}}(2)=s_{\bm{k},1} \gamma_{\bm{k},3} +s_{\bm{k},2} \gamma_{\bm{k},4} +t_{\bm{k},1} \gamma_{-\bm{k},3}^\dag+t_{\bm{k},2} \gamma_{-\bm{k},4}^\dag, \nonumber \\
&&d_{\bm{k}}(2) =s_{\bm{k},3} \gamma_{\bm{k},1} +s_{\bm{k},4} \gamma_{\bm{k},2} +t_{\bm{k},3}\gamma_{-\bm{k},1}^\dag + t_{\bm{k},4} \gamma_{-\bm{k},2}^\dag, \nonumber
\end{eqnarray}
we can obtain the diagonalized bosonic Hamiltonian as
\begin{eqnarray}
H = \sum_{\bm{k}, \alpha = 1,2,3,4} \varepsilon_{\bm{k},\alpha} \left( \gamma^\dagger_{\bm{k},\alpha} \gamma_{\bm{k},\alpha} +\frac{1}{2}\right) ,
\end{eqnarray}
where RBZ refers to the reduced Brillouin zone and $\gamma_{\bm{k},\alpha}$ stands for the Goldstone bosons of band $\alpha$. The dispersions for the 4 bands are
\begin{eqnarray}
&& \varepsilon_{\bm{k},1} = \left[  - A_{1,\bm{k}} - \left( A_{1,\bm{k}}^2 - 4B_{1,\bm{k}}\right)^{\frac{1}{2}}  \right]^{\frac{1}{2}} \\
&& \varepsilon_{\bm{k},2} = \left[  - A_{1,\bm{k}} + \left( A_{1,\bm{k}}^2 - 4B_{1,\bm{k}}\right)^{\frac{1}{2}}  \right]^{\frac{1}{2}} \\
&& \varepsilon_{\bm{k},3} = \left[  - A_{2,\bm{k}} - \left( A_{2,\bm{k}}^2 - 4B_{2,\bm{k}}\right)^{\frac{1}{2}}  \right]^{\frac{1}{2}} \\
&& \varepsilon_{\bm{k},4} = \left[ - A_{2,\bm{k}} + \left( A_{2,\bm{k}}^2 - 4B_{2,\bm{k}}\right)^{\frac{1}{2}}  \right]^{\frac{1}{2}},
\end{eqnarray}
where
\begin{eqnarray}
\nonumber  A_{1/2,\bm{k}} = && 2K_1\left( 2J_1 - K_1\right)\left| \cos\left( \frac{k_y}{2}\right)\right|^2 \\
\nonumber &&\mp 6 K_1\left( K_1 - J_1 \right) \left| \cos\left(\frac{k_y}{2}\right)\right| \\
&& + 2 K_1 \left(J_1 - 3 K_1\right),\\
\nonumber B_{1/2,\bm{k}} = && K_1^3\left( K_1 - 2J_1  \right) \left| \cos\left( \frac{k_y}{2}\right)\right|^2 \\
 \nonumber && \mp 2 K_1^2 \left( J_1^2 - K_1^2 \right) \left| \cos\left( \frac{k_y}{2}\right)\right|\\
  &&+  K_1^2 \left( K_1^2 - 2 J_1^2 + 2J_1 K_1 \right).
\end{eqnarray}
We note that the the dispersions in the AFQ23 are independent of $k_x$ in this setup (If we choose the other completely degenerate order with $\bm{q}= (\pi,2\pi/3)$ the dispersions would be independent of $k_y$), and there are a set of gapless lines for $\varepsilon_{\bm{k},3}$ along $k_x$-axis ($k_y = 0$). Due to the gapless lines, we are not able to extract the reduced order moment within the flavor-wave theory since the quantum corrections to the ordered moment are divergent in the whole regime at $K_1/J_1 \geq 1$.  

For the nematic orders, we find that $\sigma_1$ and $\sigma_Q$ are both \textit{finite} indicating that the C$_4$ lattice rotation symmetry is broken. Explicitly, we find that the nematic orders can be expressed in terms of the coefficients for the Bogoliubov transformations, as $\sigma_1 = \sum_{\bm{k}} \xi_{\bm{k},1}$ and $\sigma_Q = \sum_{\bm{k}}\xi_{\bm{k}, Q}$, with
\begin{eqnarray}
\nonumber \xi_{\bm{k},1} = && v_{\bm{k}, 1}u_{\bm{k}, 3}+v_{\bm{k},2}u_{\bm{k},4} - v_{\bm{k},1}v_{\bm{k},3} - v_{\bm{k},2}v_{\bm{k},4} + \\
\nonumber &&+t_{\bm{k},1}s_{\bm{k},3}+t_{\bm{k},2}s_{\bm{k},4}-t_{\bm{k},1}t_{\bm{k},3}-t_{\bm{k},2}t_{\bm{k},4}-\\
\nonumber  &&  - \frac{f_{\bm{k}}}{2}(v_{\bm{k},1}u_{\bm{k},1}+v_{\bm{k},2}u_{\bm{k},2}-v_{\bm{k},1}^2-v_{\bm{k},2}^2)+\\
&& + \frac{f_{\bm{k}}}{2}(t_{\bm{k},1}s_{\bm{k},1}+t_{\bm{k},2}s_{\bm{k},2}-t_{\bm{k},1}^2-t_{\bm{k},2}^2) ,
\end{eqnarray}
and
\begin{eqnarray}
\nonumber  \xi_{\bm{k},Q} = && v_{\bm{k},1}(u_{\bm{k},3} + v_{\bm{k},3} - v_{\bm{k},1})  +  \\
\nonumber && + v_{\bm{k},2}(u_{\bm{k},4} + v_{\bm{k},4} - v_{\bm{k},2}) + \\
\nonumber && + t_{\bm{k},1}(s_{\bm{k},3} + t_{\bm{k},3} - t_{\bm{k},1}) + \\
\nonumber && + t_{\bm{k},2} (s_{\bm{k},4} + t_{\bm{k},4} - t_{\bm{k},2}) + \\
\nonumber && + f_{\bm{k}} \bigg[ t_{\bm{k},1} s_{\bm{k},1} + t_{\bm{k},2} s_{\bm{k},2} -v_{\bm{k},1} u_{\bm{k},1}   -v_{\bm{k},2} u_{\bm{k},2} \bigg],\\
\end{eqnarray}
where $f_{\bm{k}} \equiv \left| e^{i k_x} + e^{-i k_y} \right|$. 
From these explicit expressions for the integrand, we find that the contributions from the gapless lines are finite. This is confirmed by numerical evaluations of these integrals: The results for $\sigma_1$ and $\sigma_Q$ are illustrated in Fig.~\ref{supp:fig_afqnem}. Thus, the nematic orders are always \textit{finite} indicating the broken C$_4$ lattice rotation symmetry in the AFQ23 order.
\end{enumerate}

\section{DMRG results}
\label{app2}

\begin{figure}
\includegraphics[width = 1.0\linewidth]{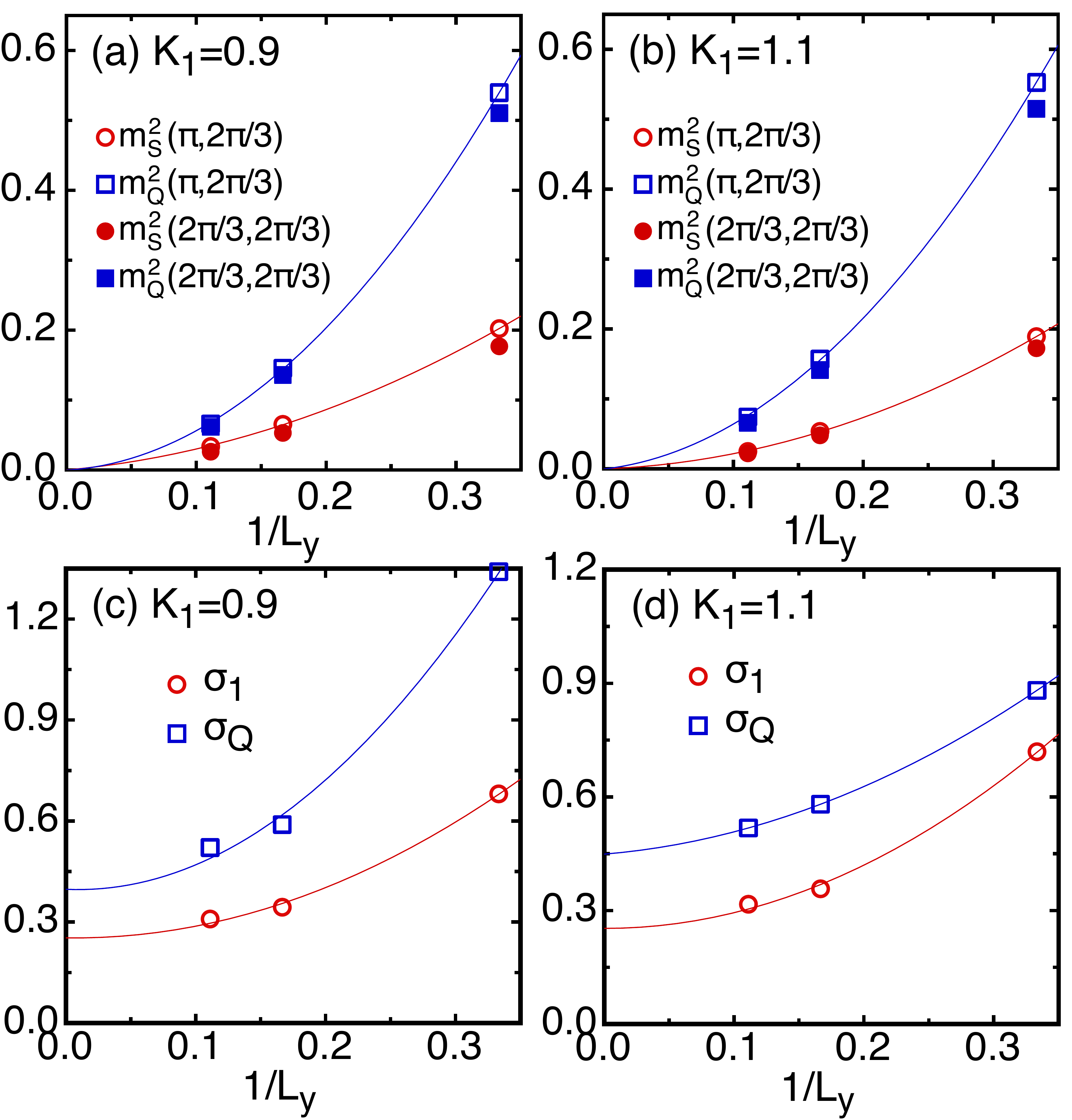}
\caption{Finite-size scaling of the spin $m^2_S$ (a) and quadrupolar $m^2_Q$ (b) order parameters for two momenta $(\pi,2\pi/3)$ and $(2\pi/3,2\pi/3)$ at $K_1=0.9$ and $1.1$. (c) and (d) are the finite-size scaling of $\sigma_1$ and $\sigma_Q$ for $K_1=0.9$ and $1.1$. The lines are guides to the eye.}\label{k09k11}
\end{figure}

We have shown the decay of magnetic and quadrupolar order parameters at the SU(3) point via finite-size scaling in Fig.~\ref{su3}(b) in the main text. Here, we complement these findings by showing the finite-size scaling of the order parameters for two different values of $K_1 = 0.9$ and $1.1$, both inside the disordered phase proposed in the present work. In Figs.~\ref{k09k11}(a)-(b), we find the spin and quadrupolar orders also scaling to zero, which are consistent with our results for the SU(3) point in the main text. In Figs.~\ref{k09k11}(c)-(d), we further show the scaling of the nematic order parameters that are defined as $\sigma_1 \equiv \frac{1}{N_m}\sum_i[\langle {\bf S}_{i}\cdot {\bf S}_{i+\hat{x}}\rangle - \langle {\bf S}_{i}\cdot {\bf S}_{i+\hat{y}}\rangle]$ and $\sigma_Q \equiv \frac{1}{N_m}\sum_i[\langle {\bf Q}_{i}\cdot {\bf Q}_{i+\hat{x}}\rangle - \langle {\bf Q}_{i}\cdot {\bf Q}_{i+\hat{y}}\rangle]$ in the bulk of cylinder ($\hat{x}$ and $\hat{y}$ denote the unit vectors along the two directions, $N_m$ is the number of sites of the two columns in the middle of cylinder). Consistent with the findings at the SU(3) point in Fig.~\ref{sigma}(d) of the main text, we find that the nematic order parameters extrapolate to finite values in the thermodynamic limit, indicating a spontaneous lattice rotational symmetry breaking in this disordered phase.

\begin{figure}
\includegraphics[width = 0.8\linewidth]{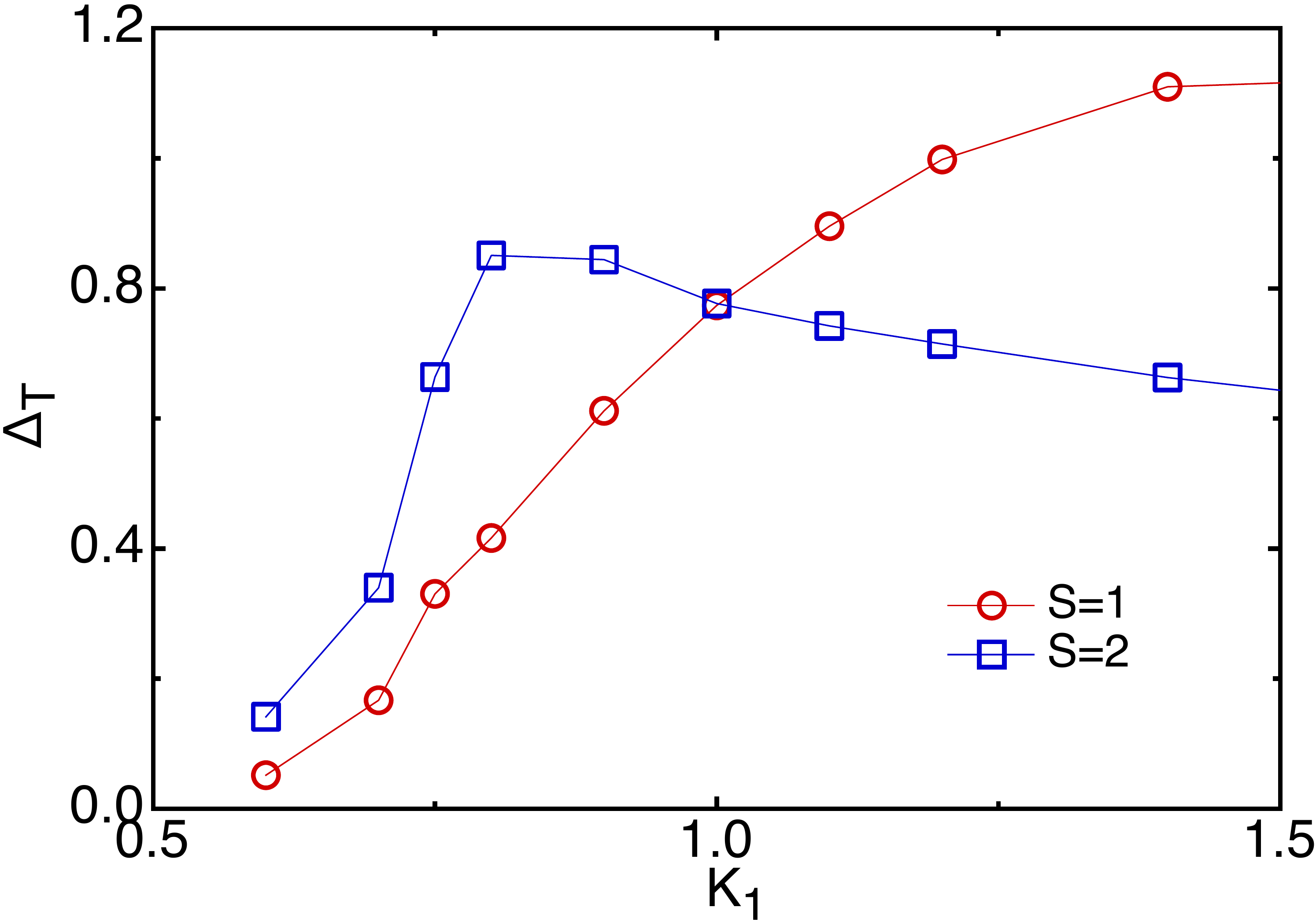}
\caption{The spin gaps $\Delta_T(S=1,2)$ as a function of $K_1$ on the RC6 cylinder.}\label{gap6}
\end{figure}

\begin{figure}
\includegraphics[width = 0.9\linewidth]{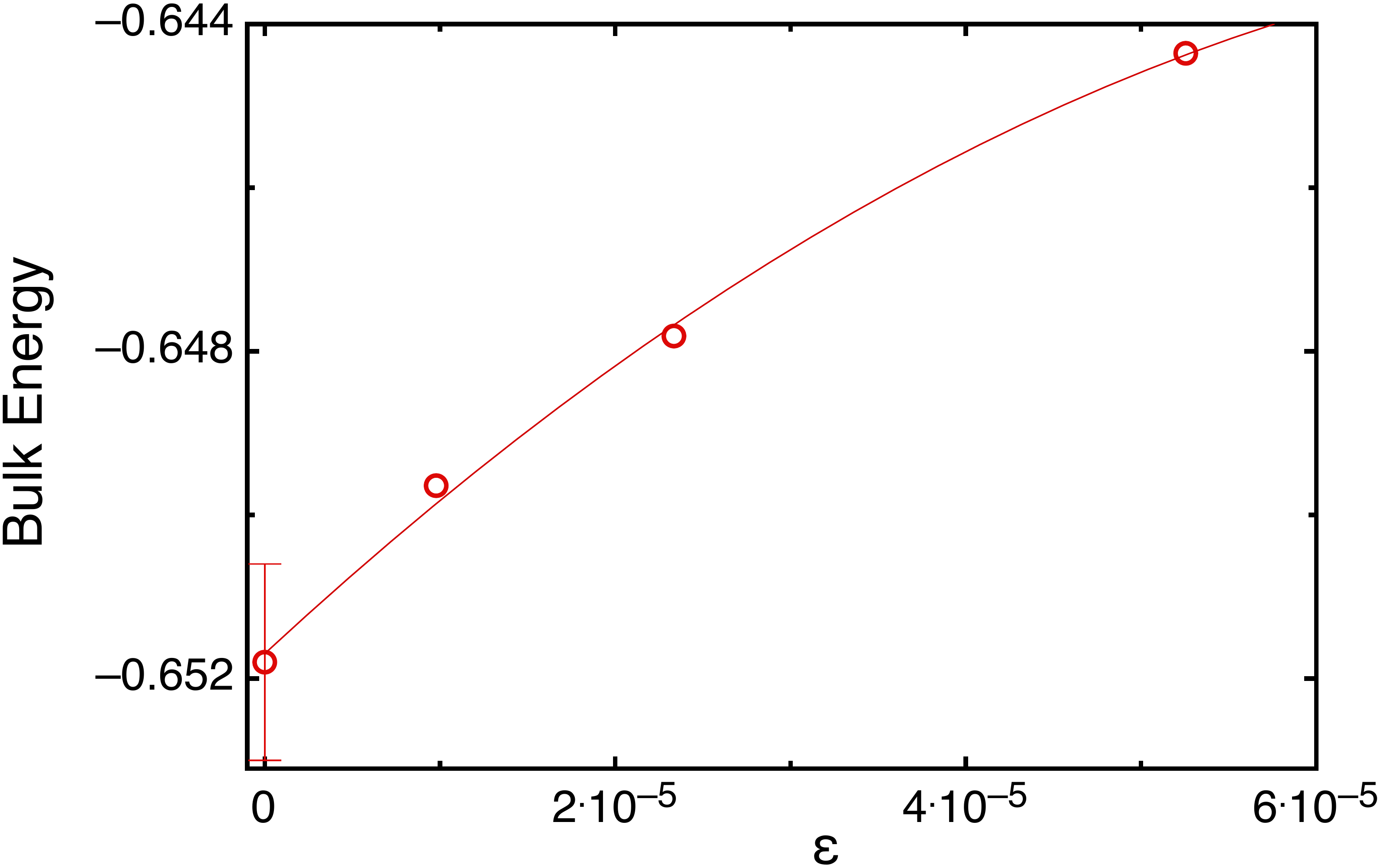}
\caption{The quadratic fitting of the bulk energy for the SU(3) model on the RC9 cylinder as a function of DMRG truncation error. We keep the SU(2) DMRG states up to $4000$.}\label{rc9e}
\end{figure}

\begin{table*}
\begin{tabular}{c|cccccc}
\hline
\hline
 & $e$ & $M_{SU(2)}$ & $\varepsilon$ & $\Delta_T(S=1)$ & $\Delta_T(S=2)$ & $\Delta_S$ \\
\hline
RC3 & $-0.8652$ & $2000$ & $4\times10^{-13}$ & $2.028$ & $2.028$ & $1.853$\\	
\hline
RC4 & $-0.6360$ & $4000$ & $4\times10^{-6}$ & $$ & $$ & $$\\
\hline
RC5 & $-0.6459$ & $4000$ & $6\times10^{-6}$ & $$ & $$ & $$\\
\hline
RC6 & $-0.6764$ & $4000$ & $2\times10^{-7}$ & $0.774$ & $0.777$ & $0.285$\\
\hline
RC7 & $-0.6392$ & $4000$ & $6\times10^{-5}$ & $$ & $$ & $$\\
\hline
RC8 & $-0.6405$ & $4000$ & $4\times10^{-5}$ & $$ & $$ & $$\\
\hline
RC9 & $-0.6518(12)$ &  &  &  &  & \\
\hline
\hline
TC3 & $-0.6825$ & $2000$ & $4\times10^{-13}$ & $$ & $$ & $$\\
\hline
TC4 & $-0.6612$ & $4000$ & $2\times10^{-6}$ & $$ & $$ & $$\\
\hline
\hline
\end{tabular}
\caption{The bulk energy per site $e$, the spin gaps $\Delta_T(S=1,2)$, and the spin singlet gap $\Delta_S$ of the SU(3) model on the RC and TC cylinders. $M_{SU(2)}$ and $\varepsilon$ are the kept SU(2) DMRG states and the DMRG truncation error, respectively. The bulk energy on the RC9 is obtained by the quadratic fitting of the energy as a function of the truncation error as shown in Fig.~\ref{rc9e}. We keep the SU(2) states up to $4000$ with the truncation error $\varepsilon \sim 1 \times 10^{-5}$. For the RC cylinder, one can see that the RC3, RC6, and RC9 cylinders have the lower energy and the smaller truncation error than the other cylinders.}\label{table}
\end{table*}

In Fig.~\ref{gap6}, we show the spin gaps between the ground state in the total spin $S = 0$ sector and the lowest-energy states in the total spin $S = 1$ and $S = 2$ sectors, respectively; which are denoted as $\Delta_{T}(S=1)$ and $\Delta_{T}(S=2)$. The gaps are obtained on the RC6 cylinder by sweeping the ground state first and then targeting the total spin $S = 1$ and $S = 2$ sectors to find the lowest-energy state. In the N\'eel AFM phase for $K_1 \lesssim 0.75$, both spin gaps are quite small, consistent with the broken spin rotational symmetry. Interestingly, at the transition $K_1 \simeq 0.75$, the spin gap $\Delta_{T}(S=2)$ undergoes a sharp increase on the RC6, while the spin-triplet gap $\Delta_{T}(S=1)$ grows gradually. For large $K_1$, the AFQ order is dominant, which agrees with the smaller spin gap $\Delta_{T}(S=2)$ we find for $K_1 > 1$. For the SU(3) model, we find that the two spin gaps have the same values on our studied systems (see also Table.~\ref{table}). In Table.~\ref{table}, we list the ground-state bulk energy and spin gaps for the SU(3) model on different geometries. The ground-state energy for RC9 is obtained by extrapolating the energy versus DMRG truncation error as shown in Fig.~\ref{rc9e}. 

\begin{figure}
\includegraphics[width = 0.9\linewidth]{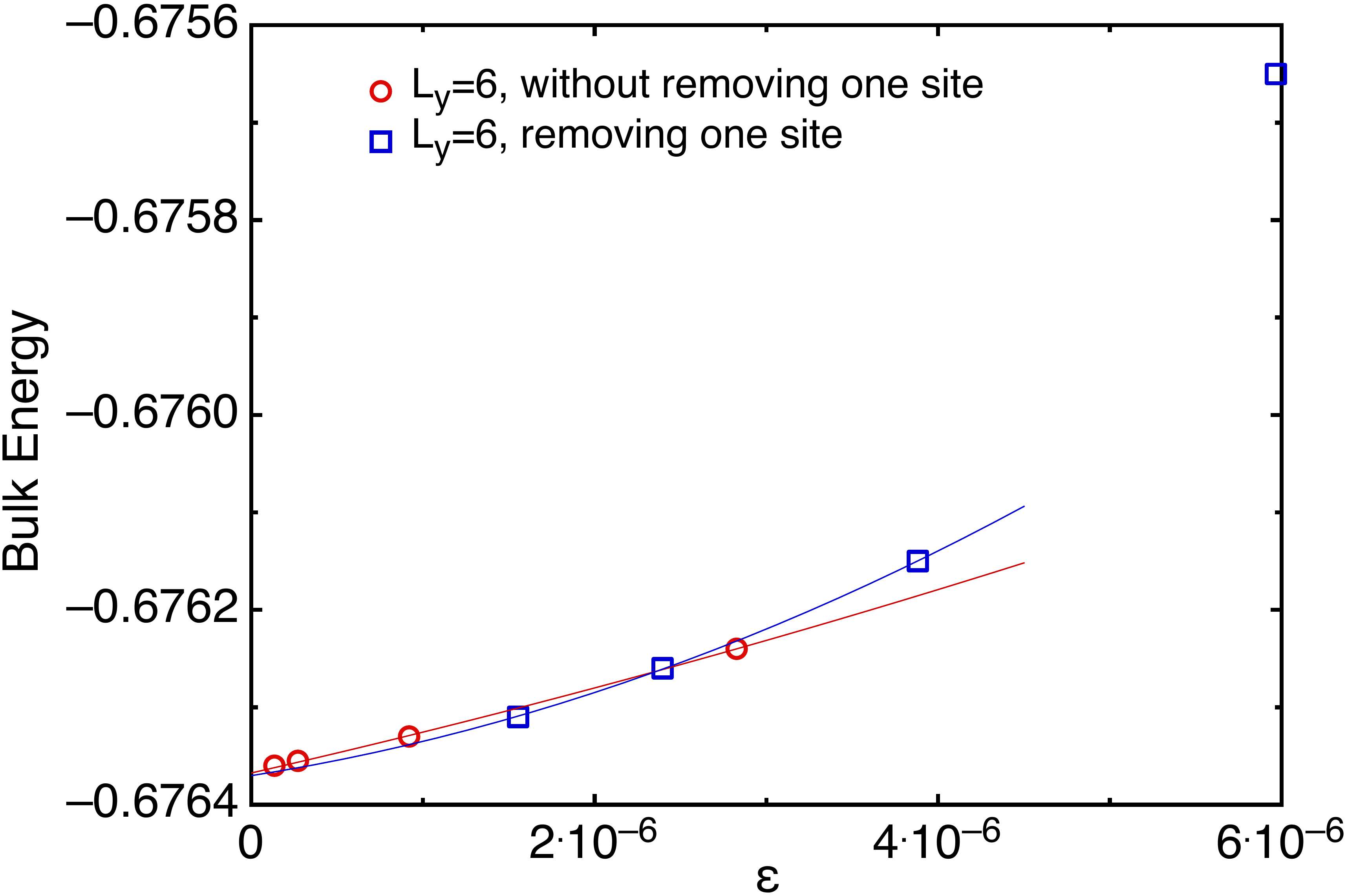}
\caption{Quadratic fitting of the bulk energy with and without removing a spin-$1$ on the RC6 cylinder as a function of DMRG truncation error. The calculations are performed by keeping the SU(2) DMRG states up to $4000$.}\label{Fig:bulk6}
\end{figure}

In DMRG simulation for a topological gapped spin liquid, the spinon sector usually can be found by removing or adding a spinon on each open edge of cylinder. On a finite-size system, the normal sector and the spinon sector have an energy difference that decays exponentially with increasing system width. This edge pinning technique has been successfully used for studying gapped spin liquids in spin-$1/2$ systems~\cite{yan2011}. For spin-$1$ gapped spin liquid, a spinon may carry spin-$1$ spin. Thus, we have performed the pinning technique for the nematic disordered phase by removing a spin-$1$ on each open edge of cylinder on the $L_y=3, 6$ cylinders. We find that for $L_y = 3, 6$, the ground states with and without removing spin-$1$ sites have the same bulk energy (see Fig.~\ref{Fig:bulk6}), which seems not to support a near-degenerate spinon sector.

\begin{figure}
\includegraphics[width = 1.0\linewidth]{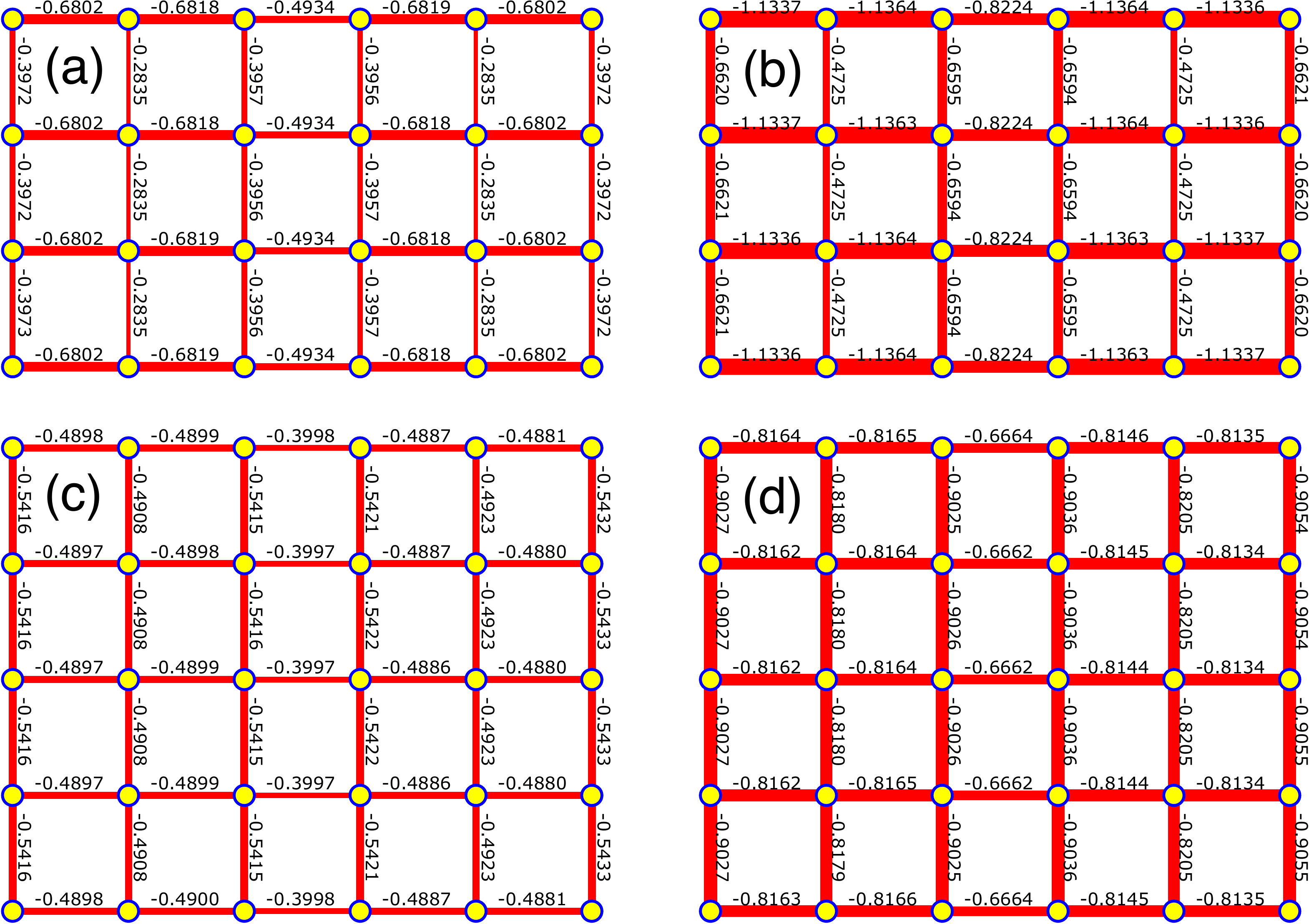}
\caption{The ${\bf S}_{i}\cdot {\bf S}_{j}$ (a) and ${\bf Q}_{i}\cdot {\bf Q}_{j}$ (b) on the nearest neighbor bonds for the SU(3) point in the middle of RC4. The ${\bf S}_{i}\cdot {\bf S}_{j}$ (c) and ${\bf Q}_{i}\cdot {\bf Q}_{j}$ (d) on the nearest neighbor bonds for the SU(3) point in the middle of RC5.}\label{sigma45}
\end{figure}

On the RC cylinder with $L_y$ not a multiple of $3$, we find the ground state with a spontaneous lattice translational symmetry breaking in the bulk of the cylinder. In Fig.~\ref{sigma45}, we show the bulk bond energy $\langle {\bf S}_{i}\cdot {\bf S}_{j}\rangle$ and $\langle {\bf Q}_{i}\cdot {\bf Q}_{j}\rangle$ on the RC4 and RC5 cylinders. In all the cases, the bond energy is not uniform and shows a period of $3$ along the $x$ direction, suggesting the breaking of translational symmetry. We note that on these geometries, the bulk energy is higher than those on the RC cylinder with $L_y = 3n$ such as RC3, RC6, and RC9, which may be owing to that the TC lattice frustrates the short-range $(\pi, 2\pi/3)$ order pattern.

\begin{figure}
\includegraphics[width = 0.8\linewidth]{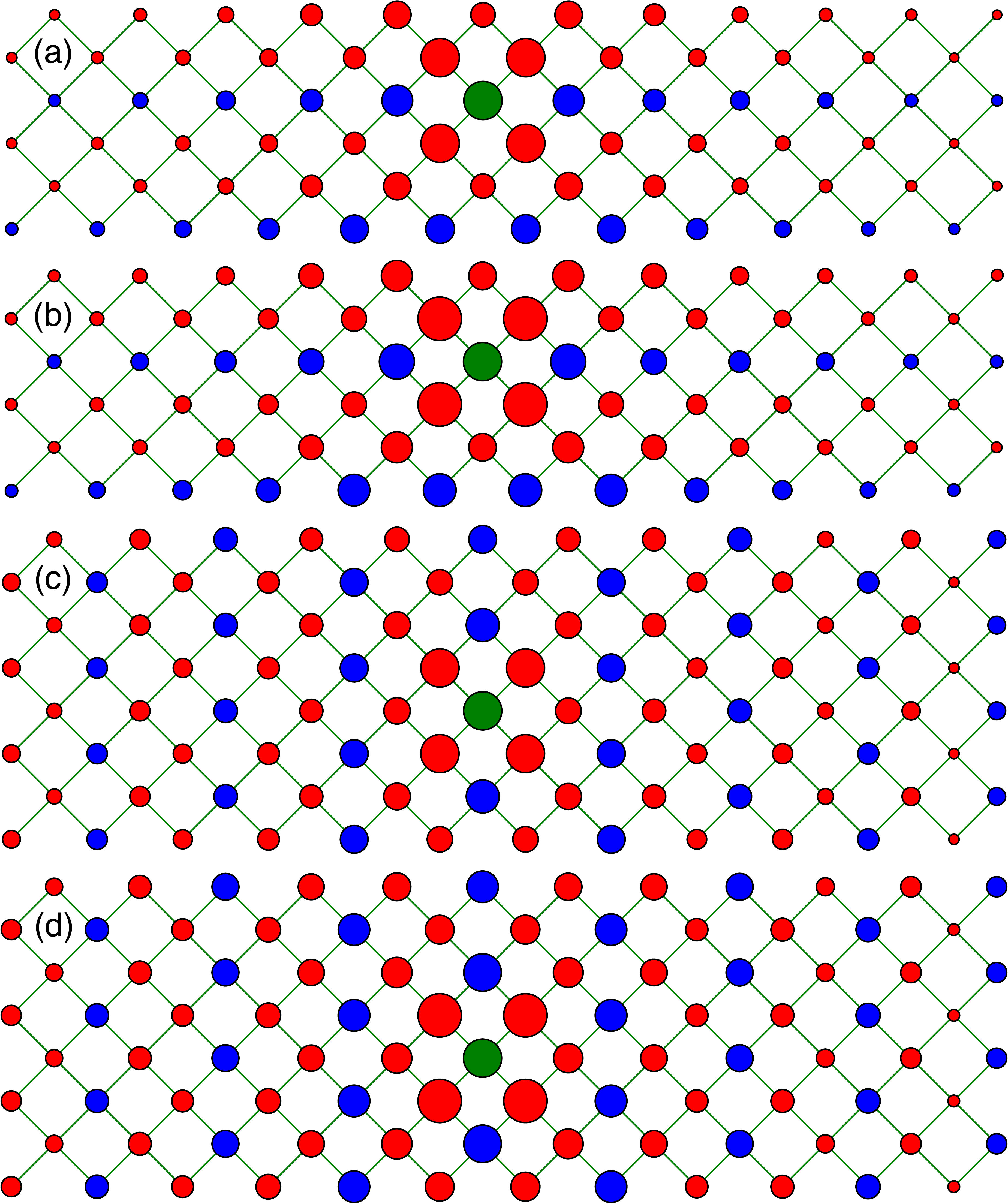}
\caption{The real space spin ((a) and (c)) and quadrupolar ((b) and (d)) correlation functions for the SU(3) model on the TC3 ((a)-(b)) and TC4 ((c)-(d)) cylinders. The green site is the reference site; the blue and red colors denote positive and negative correlations of the sites with the reference site, respectively. The area of circle is proportional to the magnitude of the spin or quadrupolar correlations.}\label{tc}
\end{figure}

On the $\pi/4$ tilted cylinder (TC), the lattice geometry frustrates the momentum at $(\pi, 2\pi/3)$ but is compatible with $(2\pi/3, 2\pi/3)$. Accordingly, we find the three-sublattice order on the TC cylinders, as shown in Fig.~\ref{tc}. DMRG calculations on the TC cylinder are harder to converge than the calculations on the RC cylinder. Here, we only show the convergent results on TC3 and TC4. We would like to mention that as shown in Table~\ref{table}, the truncation error of TC4 is much larger than the one of RC6. Consequently, we conclude that the DMRG calculations on the RC geometry are more accurate than the TC geometry at the SU(3) point, as we therefore focus on the RC geometry for the most part of this work.

\section{Spinon-Gauge Theory for Weakly-coupled SU(3) chains}
\label{app3}

As far as the origin of the $(\pi,2\pi/3)/(2\pi/3,\pi)$ orders at $L_y$-leg ladder near the SU(3) point in Figs.~2-3, which vanish at thermodynamic limit, we can also rely on the picture of the gapless spin liquid. Focusing on the SU(3) point, where the gapless disordered phase is present, we can introduce a $8$-component operator $\mathfrak{Q}^\mu$, with $\mu=1\sim 8$, consisting of $3$ components of spin $(S^{\alpha = x,y,z})$ and $5$ components of the quadrupolar moment $(Q^{i = 1\sim 5})$. The operator can be concisely expressed in terms of $3$-flavor partons, $f_{\alpha = x, y, z}$ which couple to a U(1) gauge field to gauge out the total charge mode, as $\mathfrak{Q}^\mu = f^\dagger_\alpha \left(\mathcal{M}^\mu\right)_{\alpha \beta} f_\beta$, where $M^\mu$ are $3\times 3$ matrices constructed based on Gell-Mann matrices. For a single SU(3) chain, we can expand the operator in terms of continuum field $\mathfrak{Q}^\mu \simeq \sum_{P=R/L}\mathcal{Q}^\mu_P  + \sum_{q = \pm 2 k_F} \mathcal{Q}^\mu_q e^{i q x}$, where $k_F = \pi/3$. Using Bosonization method \cite{Shankar_bosonization, Marston_bosonization, Lai_bosonization_spin1}, we find that the scaling dimensions for each are $\Delta [ \mathcal{Q}^\mu_{P=R/L } ] = 1$ and $\Delta [ \mathcal{Q}^\mu_{\pm 2\pi/3} ] = 2/3$, which leads to the power-law behaviors in the real-space correlation function, $\langle \bm{\mathfrak{D}}(x) \cdot \bm{\mathfrak{D}}(0) \rangle \sim r^{-2} + r^{-4/3} \cos(2 k_F x)$, consistent with the previous results~\cite{Itoi_Kato_spin1}. We can clearly see that the wavevector $2\pi/3$ indeed appears at a single SU(3) chain limit. We then consider the multiple SU(3) chains coupled by weak nearest-neighbor inter-chain interactions. Based on weak-coupling analysis, we find that the inter-chain interaction can be written as $H' = \sum_y \int dx \mathcal{H}'$, where
\begin{eqnarray}
\mathcal{H}' = g_1  \mathcal{Q}^\mu_{q,y}\mathcal{Q}^\mu_{\bar{q},y+1} + \lambda_{bs} \mathcal{Q}^\mu_{R,y} \mathcal{Q}^\mu_{L, y} + g_2 \mathcal{Q}^\mu_{P,y} \mathcal{Q}^\mu_{P, y+1},~~
\end{eqnarray}
where repeated indices means summation and $\bar{q} = -q = \pm 2k_F$. Based on the scaling dimension analysis, we find that only the coupling $g_1$ is relevant, which suggests that the oscillating components $\mathcal{Q}^\mu_{q,y}$ at $y$-leg and $\mathcal{Q}^\mu_{q,y+1}$ sitting at $(y+1)$-leg tend to ``anti-allign'' with each other leading to a $\pi$ period along $y$ direction. The above weak-coupling analysis for the SU(3) ladder system gives the tendency toward the formation of $(2\pi/3,\pi)/(\pi,2\pi/3)$ order, which is frustrated in the thermodynamic limit.

We now present further details to make the above points more explicit. In two dimensions (2D) the usual approach is to decompose the $3$-component spin $(S^{\alpha = x,y,z})$ and $5$-component quadrupole $(Q^{ i = 1\sim 5})$ operators in terms of $3$-flavor spinor, the fermionic partons. At SU(3) point, we can concisely construct a $8$-component operator $\mathfrak{Q} \equiv f^\dagger_\alpha M^\mu_{\alpha \beta} f_\beta$, with
\begin{eqnarray}
&& \mathfrak{Q}^{\mu = 1\sim 3} = S^\alpha = -i \epsilon^{\alpha \beta \gamma} f_\beta^\dagger f_\gamma, \\
&& \mathfrak{Q}^{\mu = 4\sim 8} =\mathcal{Q}^{i = 1\sim 5}, \\
&& f^\dagger_\alpha f_\alpha =1
\end{eqnarray}
where repeated indices mean summation, and we can identify $Q^1= Q^{x^2 - y^2}$, $Q^2 = Q^{3z^2 - r^2}$, $Q^3 = Q^{xy}$, $Q^4 = Q^{yz}$, and $Q^5 = Q^{zx}$ in the usual convention for the definition of the quadrupolar operator, and the $8$-component matrices $M$ can be related to Gell-Mann matrices $\lambda_{\mu = 1\sim8}$ as $M^1 = \lambda_7$, $M^2 = - \lambda_5$, $M^3 = \lambda_2$, $M^4 = - \lambda_3$, $M^5 = \lambda_8$, $M^6 = -\lambda_1$, $M^7 = -\lambda_6$, and $\lambda^8 = -\lambda_4$. The Gell-Mann matrices are
\begin{eqnarray}
\nonumber  && \lambda_1 = \begin{pmatrix} 
0 & 1 & 0 \\
1 & 0 & 0\\
0& 0& 0
\end{pmatrix}, ~~
\lambda_2 = \begin{pmatrix} 
0 & -i & 0 \\
i & 0 & 0\\
0& 0& 0
\end{pmatrix},~~ 
\lambda_3 = \begin{pmatrix} 
1 & 0 & 0 \\
0 & -1 & 0\\
0& 0& 0
\end{pmatrix},~\\
\nonumber &&  \lambda_4 = \begin{pmatrix} 
0 & 0 & 1 \\
0 & 0 & 0\\
1 & 0 & 0
\end{pmatrix},~~
\lambda_5 = \begin{pmatrix} 
0 & 0 & -i \\
0 & 0 & 0\\
i & 0 & 0
\end{pmatrix},~~
\lambda_6 = \begin{pmatrix} 
0 & 1 & 0 \\
0 & 0 & 1\\
0& 1 & 0
\end{pmatrix},~\\
&& \lambda_7 = \begin{pmatrix} 
0 & 0 & 0 \\
0 & 0 & -i\\
0& i & 0
\end{pmatrix},~~
\lambda_8 = \frac{1}{\sqrt{3}}\begin{pmatrix} 
1 & 0 & 0 \\
0 & 1 & 0\\
0& 0& -2
\end{pmatrix}.~
\end{eqnarray}
In the mean-field approach, one assumes that the partons are non-interacting and hopping freely on the lattice~\cite{Lee2006}. This artificially enlarges the Hilbert space, since the non-interacting parton hopping Hamiltonian allows for unoccupied and doubly-occupied sites, which are strictly forbidden in the present quantum spin model. One route to project the enlarged Hilbert space into a physical one is to perform Gutzwiller projection to project the enlarged Hilbert space at the mean field level back into the physical Hilbert space for the quantum spin model restricting the patrons to single occupancy. The alternate approach for implementing the constraint of the single occupancy is by introducing a gauge field, for which the simplest case is the U(1) gauge field, minimally coupled to the patrons in the hopping Hamiltonian. By doing this, the theory becomes a strongly-coupled lattice gauge field theory, which is hard to be solved analytically. Fortunately, on the chain limit, we can employ Bosonization to analyze the quasi-1D gauge theory, which can capture universal low energy properties of the ground state in the spin Hamiltonian~\cite{Lee2006,sheng2009}.

We now start by using Bosonization to analyze the gauge theory. We assume a mean field state in which the partons are hopping on the chain with nearest-neighbor hopping strengths denoted $t_1$. The dispersion of each flavor of parton is
\begin{eqnarray}
\xi(k) = - 2 t_1 \cos (k).
\end{eqnarray}
At the mean-field level, each flavor of parton is $1/3$ filled, which gives one set of Fermi crossings at wave vectors $\pm k_F = \pm \pi/3$. The parton operators are expanded in terms of continuum fields,
\begin{eqnarray}
f_\alpha(x) = \sum_{P} e^{i P k_F x} f_{P\alpha},
\end{eqnarray}
with $\alpha = x, y, z$ denoting the flavor, and $P = R/L = \pm$ denoting the right and left mov-ing fermions. We  now use Bosonization re-expressing these low energy parton operators with Bosonic fields,
\begin{eqnarray}
f_{P\alpha} = \eta_{\alpha} e^{i \left( \varphi_{\alpha} + P \theta_{\alpha}\right)},
\end{eqnarray}
with canonically conjugate boson fields:
\begin{eqnarray}
&& [ \varphi_\alpha (x), \varphi_\beta (x')] = [ \theta_\alpha(x), \theta_\beta(x')]=0,\\
&& [ \varphi_\alpha (x), \theta_\beta (x')] = i \pi \delta_{\alpha \beta} \Theta(x-x'),
\end{eqnarray}
where $\Theta(x)$ is the Heaviside step function, and $\eta_\alpha$ are the Klein factors, the Majorana fermions $\{ \eta_\alpha, \eta_\beta\}$, for assuring the anti-commutation between partons with different flavors. Under bosonization, the slowly varying fermionic densities are simply $f^\dagger_{P\alpha} f_{P\alpha} = \partial_x (P\varphi_\alpha + \theta_\alpha)/(2\pi)$.

In the present 1+1D continuum theory, we work under the gauge constraint that eliminate spatial components of the gauge field. In the imaginary-time formalisim, the bosonized Lagrangian density is,
\begin{eqnarray}
\mathcal{L} = \frac{1}{2\pi}\sum_{\alpha = x,y,z} \left[ \frac{1}{v_\alpha} \left( \partial_\tau \theta_\alpha \right)^2 + v_\alpha \left( \partial_x \theta_\alpha \right)^2 \right] + \mathcal{L}_A.
\end{eqnarray}
Here $\mathcal{L}_A$ encodes the coupling to the slowly varying 1D (scalar) potential field $A(x)$,
\begin{eqnarray}
\mathcal{L}_A = \frac{1}{m} \left( \frac{\partial_x A}{\pi}\right)^2 + i \rho_A A,
\end{eqnarray}
where $\rho_A$ denotes the total gauge charge density, 
\begin{eqnarray}
\rho_A = \frac{1}{\pi}\sum_\alpha\partial_x \theta_\alpha.
\end{eqnarray}
In the present SU(3) chain, it is useful to define fields as
\begin{eqnarray}
&& \theta_\rho = \frac{1}{\sqrt{3}} \sum_{\alpha} \theta_\alpha, \\
&& \theta_1 = \frac{1}{\sqrt{2}} \left( \theta_x - \theta_y \right), \\
&& \theta_2 = \frac{1}{\sqrt{6}} \left( \theta_x + \theta_y - 2\theta_z \right),
\end{eqnarray}
and similar expressions for $\varphi_\rho,~\varphi_1,~\varphi_2$ fields, which leads to the Lagrangian density in the same form as before,
\begin{eqnarray}\label{supp:bosonic_L}
\mathcal{L} = \frac{1}{2\pi}\sum_{\alpha = \rho,1,2} \left[ \frac{1}{v_\alpha} \left( \partial_\tau \theta_\alpha \right)^2 + v_\alpha \left( \partial_x \theta_\alpha \right)^2 \right] + \mathcal{L}_A.
\end{eqnarray}
Integration over the gauge potential generate a mass term,
\begin{eqnarray}\label{supp:bosonic_gauge_constraint}
\mathcal{L}_A \sim m \left( \theta_\rho - \theta_\rho^{(0)}\right)^2,
\end{eqnarray}
for the field $\theta_\rho= \sum_\alpha \theta_\alpha/2$. Due to the presence of the mass term for the total charge mode, $\theta_\rho$, the $\theta_\rho$ becomes gapped and can be ignored essentially.

The spin and quadrupolar operators can also be re-expressed in terms of the bosonic fields, and their corresponding correlation functions can be determined based on the bosonic Lagrangian above, Eqs.~\eqref{supp:bosonic_L}-\eqref{supp:bosonic_gauge_constraint}. Let's take the spin operator as an illustration. We find that the spin operator at the low-energy theory description consists of a uniform and a oscillating parts with wave vectors $q = \pm 2k_F = \pm 2\pi/3$,
\begin{eqnarray}
S^\alpha(x) \simeq S^\alpha_{uni} (x) + \sum_{q = \pm 2k_F}S^\alpha_{q}(x) e^{i q x},
\end{eqnarray}
with
\begin{eqnarray}
&& S^\alpha_{uni}(x) \simeq \frac{1}{2\pi} \left( \partial_x \varphi_{R\alpha} - \partial_x\varphi_{L\alpha}\right) = \frac{1}{\pi} \partial_x \theta_\alpha,\\
&& S^\alpha_{q}(x) \simeq -i \epsilon^{\alpha \beta \gamma} \eta_\beta \eta_\gamma e^{-i (\theta_\beta + \theta_\gamma)} \cos\left(\varphi_\beta - \varphi_\gamma \right).
\end{eqnarray}
Based on the scaling dimension analysis, we find the scaling dimensions for the uniform and oscillating parts to be
\begin{eqnarray}
 \Delta[S^\alpha_{uni}] = 1, & ~~~~~~\Delta[S^\alpha_q] = 2/3,
\end{eqnarray}
which leads to the conclusion that a spin correlation function at such a SU(3) spin chain shows the power-law behavior as
\begin{eqnarray}
\left\langle \vec{S}(x) \cdot \vec{S}(0)\right\rangle &\simeq& r^{-2 \Delta[S^\alpha_{uni}]} + r^{-2 \Delta[S^\alpha_q]}\cos(2 k_F x) \\
& = & r^{-2} + r^{-4/3} \cos(2 k_F x),
\end{eqnarray}
which is consistent with the previous studies \cite{Itoi_Kato_spin1}. Similarly, we can find that the quadrupolar operator has the similar behaviors as
\begin{eqnarray}
\mathcal{Q}^i (x) \simeq \mathcal{Q}^i_{uni} (x) + \sum_{q=\pm 2k_F} \mathcal{Q}^i_{q} e^{i q x},
\end{eqnarray}
and their scaling dimensions are 
\begin{eqnarray}
\Delta[\mathcal{Q}^i_{uni}] = 1, & ~~~~~~\Delta[\mathcal{Q}^i_q] = 2/3.
\end{eqnarray}
We can clearly see the $2\pi/3$ wave vector. 

Now in order to see the $\pi$ wave vector found in the DMRG calculations, we proceed to consider weak couplings between each SU(3) chains. Introducing a more compact $8$-component operator $\mathfrak{Q}$ consisting both spin and quadrupolar operators, we can then expand the $8$-component in terms of low-energy fields as
\begin{eqnarray}
\mathfrak{Q}^{\mu=1\sim8} \simeq  \mathcal{Q}^\mu_{uni} + \sum_{q = \pm 2k_F} \mathcal{Q}^\mu_{q}e^{i q x}.
\end{eqnarray}
At weak-coupling regime, we find that the nearest-neighbor couplings between $y$-leg and $(y+1)$-leg chains can be written down at the low-energy description as $H' = \sum_y \int dx \mathcal{H}'$ with
\begin{eqnarray}
\mathcal{H}' = g_1  \mathcal{Q}^\mu_{q,y}\mathcal{Q}^\mu_{\bar{q},y+1} + \lambda_{bs} \mathcal{Q}^\mu_{R,y} \mathcal{Q}^\mu_{L, y} + g_2 \mathcal{Q}^\mu_{P,y} \mathcal{Q}^\mu_{P, y+1},~~
\end{eqnarray}
where the repeated indices mean summations. We define $\mathcal{Q}^\mu_{y} \equiv \sum_P \mathcal{Q}^\mu_{P,y}$, and $\bar{q} = -q$. At tree-level renormalization group (RG) analysis, we find the scaling dimensions for the couplings as $\Delta[g_1] = 2/3$, where $\lambda_{bs}$ and $g_2$ remain \textit{marginal}. Therefore, we can see if $g_1 >0$, under RG the $\mathcal{Q}^\mu_{q,y}$ and $\mathcal{Q}^\mu_{\bar{q},y+1}$ tend to ``anti-align'' with each other leading to a staggered pattern along $y$-direction with a period of $\pi$. 

We can then see based on the weak-coupling analysis of multiple SU(3) chains weakly coupled by weak inter-chain interactions, the wave vector $(2\pi/3, \pi)/(\pi,2\pi/3)$ can naturally arise, which is consistent with the DMRG results.

\bibliography{jkmodel}
\end{document}